\documentclass[a4paper,11pt]{article}
\pdfoutput=1 

\usepackage{jheppub} 
\usepackage{graphicx}
\usepackage{csquotes}
\usepackage{latexsym}
\usepackage{bbm}
\usepackage[mathscr]{euscript}
\usepackage{subcaption}
\usepackage{amsmath,amssymb,bm} 
\newcommand{\beq}{\begin{equation}}
\newcommand{\eeq}{\end{equation}}
\newcommand{\ba}{\begin{array}{ccc}}

\newcommand{\sgn}{\textrm{sgn}}
\def\bea{\begin{eqnarray}}
\def\eea{\end{eqnarray}}
\usepackage[colorlinks=true]{hyperref} 
\hypersetup{
    unicode=false,          
    pdftoolbar=true,        
    pdfmenubar=true,        
    pdffitwindow=false,     
    pdfstartview={FitH},    
    pdfsubject={},   
    pdfcreator={},   
    pdfproducer={}, 
    pdfkeywords={} {} {}, 
    pdfnewwindow=true,      
    colorlinks=true,       
    linkcolor=magenta, 
    citecolor=blue,        
    filecolor=magenta,      
    urlcolor=blue           
}

\newcommand{\psibar}{\bar{\psi}}
\renewcommand{\approx}{\simeq}

\renewcommand{\approx}{\simeq}

\renewcommand{\approx}{\simeq}

\usepackage{xspace}
\usepackage[all]{xy}
\usepackage{amsmath}
\usepackage{float}

\title{Four-point function of the complex Sachdev-Ye-Kitaev model at finite chemical potential}
\author[\mathscr{E}]{Can Onur Akyuz,}
\author[\mathscr{E}]{Erick Arguello Cruz,}
\author[\mathscr{E}]{Ludo Fraser-Taliente,}
\author[\mathscr{E}]{Grigory Tarnopolsky}
\affiliation[\mathscr{E}]{Department of Physics, Carnegie Mellon University, Pittsburgh, PA 15213, USA}

\emailAdd{cakyuz@andrew.cmu.edu}
\emailAdd{earguell@andrew.cmu.edu}
\emailAdd{lfrasert@andrew.cmu.edu}
\emailAdd{gtarnopo@andrew.cmu.edu}

\abstract{It is known that, for a range of chemical potentials, the infrared behavior of the complex Sachdev-Ye-Kitaev (cSYK) model is governed by a 1D Nearly Conformal Field Theory (NCFT$_{1}$), thereby realizing a continuous line of NCFTs. A finite chemical potential $\mu$ introduces an asymmetry parameter $\mathscr{E}$ into the cSYK fermion two-point function in the conformal limit. In this work, we compute the cSYK four-point function in the conformal limit for an arbitrary value of $\mathscr{E}$ at leading order in $1/N$. We show that the result is fully consistent with the NCFT$_{1}$ structure of the cSYK model and use it to extract the structure constants for correlation functions of two complex fermions with bilinear operators.
}

\begin{document}
\maketitle
\flushbottom

\section{Introduction}

The Sachdev-Ye-Kitaev (SYK) models \cite{SachdevYe_1993, Kitaev_talks}, together with their tensor-model counterparts \cite{Gurau:2011xq, Witten:2016iux, Klebanov:2016xxf, Giombi:2017dtl}, have become widely used theoretical laboratories for exploring strongly coupled phenomena in condensed matter physics, holography in high-energy theory, and quantum simulations (for reviews, see \cite{Sarosi:2017ykf, Rosenhaus:2018dtp,Klebanov:2018fzb, Trunin:2020vwy, Chowdhury_2022}). 
The Majorana SYK model with the Hamiltonian 
\begin{align}
H =i^{q/2}\sum_{1\leq i_1<  \cdots < i_q\leq N } J_{i_1\cdots i_q} \psi_{i_1}\cdots \psi_{i_q}    \,,
\end{align}
with all-to-all random couplings $J_{i_1\cdots i_q}$ and Majorana fermions $\psi_{i}$ provides one of the cleanest and most tractable examples \cite{Maldacena:2016hyu, Polchinski:2016xgd, Cotler:2016fpe, Jevicki:2016bwu, Jevicki:2016ito, Garcia-Garcia:2016mno, Gross:2017aos, Bagrets:2017pwq, Stanford:2017thb, Mertens:2017mtv, Kitaev:2017awl, Garcia-Garcia:2017pzl}. 
While the experimental detection and controlled manipulation of Majorana zero modes remain major challenges for modern experiments, the complex SYK model \cite{Sachdev:2015efa, Davison:2016ngz} provides a more realistic alternative\footnote{Similar random models were proposed in \cite{FRENCH1970449, BOHIGAS1971261} for describing energy levels of a nucleus. These nuclear shell models assumed a fixed number of nucleons with a fixed total angular momentum.}. There are various proposals for simulating this model \cite{Danshita:2016xbo, Pikulin:2017mhj, Chen_2018, Wei_2021, Wei_2023}. Besides being more amenable to experimental quantum simulation at present, the complex SYK model also features an additional control parameter—the chemical potential—and exhibits new, interesting properties compared with the Majorana SYK model.

The Hamiltonian of the complex Sachdev-Ye-Kitaev model (cSYK)  reads   \cite{Sachdev:2015efa, Davison:2016ngz, Gu:2019jub}
\begin{align}
H =\sum_{\substack{  1\leq i_1<  \ldots < i_{q/2}\leq N \\ 1\leq j_1<  \ldots < j_{q/2}\leq N}}
J^{i_1\cdots i_{q/2}}_{j_1\cdots j_{q/2}} \, \mathcal{A}\{\bar{\psi}_{i_1}\cdots \bar{\psi}_{i_{q/2}} \psi^{j_1}\cdots \psi^{j_{q/2}}\}  - \mu \sum_{i} \bar{\psi}_{i} \psi^{i}\,,   
\label{complex_Hamiltonian}
\end{align}
where $\psi^{i}, \bar{\psi}_{i}=(\psi^i)^\dag$, with $i=1,\dots, N$ are $N$ complex fermions,  which interact with each other through random complex couplings $J_{i_{1}\dots i_{q/2}}^{j_{1}\dots j_{q/2}}$ with $q$ an even integer number and $\mu$ the chemical potential. 
The symbol $\mathcal{A}\left\lbrace \dots \right\rbrace$ denotes the anti-symmetrized product of operators, which imposes an exact particle-hole symmetry $\psi^{i} \leftrightarrow \bar{\psi}_{i}$ of the Hamiltonian at $\mu=0$.  The complex fermions $\bar{\psi}_i$ and $\psi^i$ satisfy 
$\{\psi^i,\bar{\psi}_j\}=\delta^i_j$,
$\{\psi^i,\psi^j\}=\{\bar{\psi}_i,\bar{\psi}_j\}=0$ with
$i,j=1,2,\ldots,  N$. 
The couplings $J^{i_1\cdots i_{q/2}}_{j_1\cdots j_{q/2}}$ are Gaussian complex random variables with zero mean and  variance 
\begin{align}
\left\langle J^{i_1\cdots i_{q/2}}_{j_1\cdots j_{q/2}}
J_{i_1\cdots i_{q/2}}^{j_1\cdots j_{q/2}}\right\rangle
=J^2\frac{(q/2)!(q/2-1)!}{N^{q-1}}\,, \label{JJvar}
\end{align}
and also satisfy
$(J^{i_1\cdots i_{q/2}}_{j_1\cdots j_{q/2}})^{*}=J_{i_1\cdots i_{q/2}}^{j_1\cdots j_{q/2}}$  in order for the Hamiltonian to be Hermitian.   
The conserved $U(1)$ charge density operator $\hat{Q}$ is defined as
\begin{align}
\hat{Q} \equiv  \frac{1}{N} \sum_{i} \bar{\psi}_{i}\psi^{i} -\frac{1}{2}\,,
\end{align}
and it is related to the asymmetry of the imaginary time Green's function at finite temperature 
\begin{align}
G(\tau)\equiv\frac{1}{N}\sum_{i}\langle \textrm{T} \psi^{i}(\tau) \bar{\psi}_{i}(0)\rangle, \quad G(0^{+}) = \frac{1}{2}-Q, \quad G(\beta^{-})=\frac{1}{2}+Q\,, \label{GandQ}
\end{align}
where $Q \equiv \langle \hat{Q} \rangle$ and $\beta=1/T$ is the inverse temperature. The Green's function satisfies the Kubo-Martin-Schwinger (KMS) condition $G(\tau +\beta)=-G(\tau)$ and in the large $N$ limit it obeys the Schwinger-Dyson (SD) equations \cite{Sachdev:2015efa, Davison:2016ngz}:
\begin{equation}
\begin{aligned}
&(\partial_{\tau}-\mu)G(\tau) + \int_{0}^{\beta} d\tau' \Sigma(\tau -\tau')G(\tau') = \delta(\tau)\,, \quad \Sigma(\tau) = J^{2}G(\tau)^{\frac{q}{2}}G(\beta -\tau)^{\frac{q}{2}-1}\,. \label{SDequations0}
\end{aligned}
\end{equation}
In the strong-coupling regime  $\beta J \gg 1$ (conformal limit), and for $\tau \in [-\beta,\beta]$ away from $\tau=0,\pm\beta$, the solution of \eqref{SDequations0} is approximated by the conformal two-point function \cite{Sachdev:2015efa, Davison:2016ngz}:
\begin{align}
G_{c}(\tau) =  b^{\Delta}\,\sgn (\tau) \Big|\frac{\beta J}{\pi} \sin \frac{\pi \tau}{\beta}\Big|^{-2\Delta} 
e^{2\pi \mathscr{E}(\frac{|\tau|}{\beta} - \frac{1}{2})\sgn(\tau)}\,,  \label{GcSol}
 \end{align}
with $\Delta=1/q$ and the constant 
\begin{align}
    b&= \frac{(1-2\Delta)\sin{(2\pi\Delta)}}{4\pi\cos{(\pi\Delta+i\pi \mathscr{E})}\cos{(\pi\Delta-i\pi \mathscr{E})}}= \frac{(1-2\Delta)\sin{(\pi\Delta+\theta)}\sin{(\pi\Delta-\theta)}}{\pi \sin{2(\pi\Delta)}}\,, \label{Constb}
\end{align}
where $\mathscr{E}$ and $\theta$ are called \textit{asymmetry parameters}, and are related to each other by 
\begin{align}
\label{eq_parameters}
    e^{-2i\theta}=\frac{\cos{(\pi\Delta+i \pi\mathscr{E})}}{\cos{(\pi\Delta-i \pi\mathscr{E})}},\quad e^{2\pi \mathscr{E}}=\frac{\sin{(\pi\Delta+\theta)}}{\sin{(\pi\Delta-\theta)}}\,.
\end{align}
The asymmetry parameter $\mathscr{E}$ (and $\theta$) is a new feature of the cSYK model that is absent in the Majorana SYK model. Both the $U(1)$ charge density $Q$ and the asymmetry parameter $\mathscr{E}$ (or $\theta$) depend on the chemical potential $\mu$ and also $\mathscr{E}\geq 0$ ($\theta \geq 0$) for $\mu\geq 0$ \footnote{The particle-hole symmetry maps the Hamiltonian with $+\mu$ to that with $-\mu$, therefore, without loss of generality, we can restrict the chemical potential (and thus the asymmetry parameters) to the non-negative domain.}. This dependence is not known analytically, but it has been computed numerically in \cite{Azeyanagi_2018,  Ferrari:2019ogc, Tikhanovskaya:2020elb}. Nevertheless, even though the $\mu$-dependence of $\mathscr{E}$ and $Q$ is not known analytically, one can still derive the Luttinger relation between $Q$ and $\mathscr{E}$ (or $\theta$) \cite{Georges_2001, Gu:2019jub, Tikhanovskaya:2020elb}:
\begin{equation}
\begin{aligned}
Q &= \frac{1}{2\pi i} \ln \left(\frac{\cos(\pi \Delta - i \pi \mathscr{E})}{\cos(\pi \Delta+ i \pi \mathscr{E})}\right) + \frac{1-2\Delta}{4i}\Big(\tan (\pi \Delta+i\pi \mathscr{E})-\tan(\pi \Delta-i\pi \mathscr{E})\Big) \\
&=\frac{\theta}{\pi} + \Big(\frac{1}{2}-\Delta\Big)\frac{\sin(2\theta)}{\sin(2\pi \Delta)}\,.\label{LutRel} 
\end{aligned}
\end{equation}
For the spectral density of the cSYK model to be positive, the asymmetry parameter $\theta$ must lie in the range $\theta \in (-\pi \Delta, \pi \Delta)$, which corresponds to $\mathscr{E} \in (-\infty, +\infty)$. Although the full range of the asymmetry parameters is formally allowed by the conformal solution (\ref{GcSol}), it was shown numerically in \cite{Azeyanagi_2018, Ferrari:2019ogc} that the cSYK model undergoes a first-order phase transition to a \enquote{low-entropy} phase if the absolute value of the chemical potential becomes sufficiently large. This implies that the range of $\theta$ (and $\mathscr{E}$), for which the conformal (\enquote{high-entropy}) phase is stable, reduces. For $q=4$ at zero temperature, the critical value was computed numerically in \cite{Tikhanovskaya:2020elb} to be $|\theta_{\textrm{crit}}|\approx 0.153\pi$ ($|\mathscr{E}_{\textrm{crit}}|\approx 0.184$), which corresponds to $|Q_{\textrm{max}}|\approx 0.358$.

It has been proposed that, in the large-$N$ limit and for $\beta J \gg 1$, the cSYK model is governed by a one-dimensional Nearly Conformal Field Theory (NCFT$_1$) for any value of the chemical potential $\mu$ within the conformal phase \cite{Gu:2019jub, Tikhanovskaya:2020elb}. In this sense, varying $\mu$ generates a continuous family (a line) of NCFTs.
This NCFT$_{1}$ contains an infinite set of bilinear operators, which can be schematically written as $\mathcal{O}_{m}(\tau) = \bar{\psi}_{i}\partial_{\tau}^{m}\psi^{i}$ for $m=0,1,2,\dots$. 
The scaling dimensions $h_{m}$ of these operators depend on the asymmetry parameter $\mathscr{E}$ and are obtained from solving the transcendental equations, which we write  in  (\ref{AnDims}) of  Section~\ref{4ptAsBlocks}. The special operators $\mathcal{O}_{0}(\tau)$ and $\mathcal{O}_{1}(\tau)$ correspond to the conserved $U(1)$ charge density $\hat{Q}$ and the Hamiltonian $H$, and have scaling dimensions $h_{0}=1$ and $h_{1}=2$. Although these operators are assigned integer scaling dimensions, their correlation functions do not obey the CFT power-law structure. They are thus responsible for reducing the conformal behavior to nearly conformal behavior \cite{Maldacena:2016hyu, Kitaev:2017awl}. The remaining  bilinear operators $\mathcal{O}_{2}, \mathcal{O}_{3}, \mathcal{O}_{4},\dots$ are expected to obey the usual CFT behavior. However, numerical evidence for this is currently limited to the Majorana SYK model \cite{Cruz:2022uic}.

The consistency of the cSYK description as an NCFT$_{1}$ can be tested by a direct computation of the cSYK four-point function of fermions. The result can then be compared with the predictions that follow from the operator product expansion (OPE) of the fermionic fields and the general CFT structure of correlation functions of local operators. The direct computation of the cSYK four-point function for an arbitrary asymmetry parameter $\mathscr{E}$, and its subsequent comparison with the CFT prediction, is the main result of this article.

The paper is organized as follows. In Section \ref{SeccSYK4ptfunc}, we derive a general expression for the cSYK four-point function in the large-$N$ limit by resumming the leading ladder diagrams. In Section \ref{SecConfLim}, we obtain the large-$N$ four-point function in the conformal limit. In Section \ref{4ptAsBlocks}, we write the conformal four-point function as a sum over conformal blocks of the bilinear operators. In Section~\ref{4ptCFTandStr}, we compute the four-point function using CFT analysis, and by comparing this result with the direct computation in the previous section, we infer the CFT structure constants for correlation functions of two complex fermions with bilinear operators. Several appendices provide additional details of the computations.

\section{Four-point function in the Complex SYK model}
\label{SeccSYK4ptfunc}

\begin{figure}[ht]
               \centering
              \includegraphics[width=14cm]{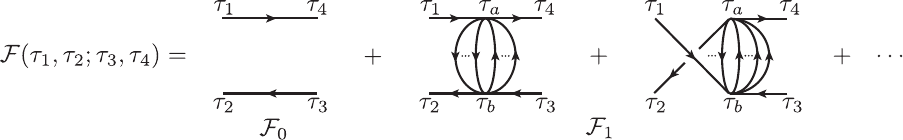}
              \caption{Ladder diagrams contributing to $\mathcal{F}(\tau_{1},\tau_{2};\tau_{3}, \tau_{4})$ can be grouped into $\mathcal{F}_n$, where $n$ labels the number of \enquote{rungs}. Each line with an arrow represents the large $N$ Green’s function $G(\tau)$, which satisfies the SD equations \eqref{SDequations0}.}
              \label{4ptfunct}
\end{figure}

The large $N$ expansion of the imaginary time four-point function at finite temperature is
\begin{align}\label{eq:Fdef}
\frac{1}{N^{2}}\langle \psi^{i}(\tau_{1}) \psibar_{i}(\tau_{2})\psi^{j}(\tau_{3})\psibar_{j}(\tau_{4})\rangle  = G(\tau_{12})G(\tau_{34}) + \frac{1}{N}\mathcal{F}(\tau_{1},\tau_{2};\tau_{3},\tau_{4})+\cdots\,,
\end{align}
where the sum over indices $i$ and $j$ is assumed and the Green's function $G(\tau)$ is obtained from the SD equations (\ref{SDequations0}).
We will compute the leading connected term $\mathcal{F}$ in this expression, which we will refer to as the four-point function below.
Resumming the melonic diagrams, we get a (geometric) series in terms of the full propagators $\mathcal{F}=\sum_{n}\mathcal{F}_{n}$, see Fig. \ref{4ptfunct}.

\noindent The first diagram $\mathcal{F}_{0}$, which is the seed for the rest, is given by the expression
\begin{align}
\mathcal{F}_{0}(\tau_{1},\tau_{2};\tau_{3},\tau_{4}) = -G(\tau_{14})G(\tau_{32})\,.
\end{align}
The term $\mathcal{F}_{n+1}$ is obtained  by applying the kernel $K$  to $\mathcal{F}_{n}$:
\begin{align}
\mathcal{F}_{n+1} (\tau_{1},\tau_{2};\tau_{3},\tau_{4}) = \int_{0}^{\beta} d\tau_{a} d\tau_{b} K(\tau_{1},\tau_{2};\tau_{a},\tau_{b}) \mathcal{F}_{n} (\tau_{a},\tau_{b};\tau_{3},\tau_{4})\,,
 \end{align}
and the kernel is not symmetric and reads  
\begin{equation}
\begin{aligned}
K(\tau_{1},\tau_{2};\tau_{3},\tau_{4})  =& (-1)^{\frac{q}{2}-1}J^{2} \Big(  \frac{q}{2} G(\tau_{13})G(\tau_{42})G(\tau_{34})^{\frac{q}{2}-1}G(\tau_{43})^{\frac{q}{2}-1} +  \\
&+\big(\frac{q}{2}-1\big)G(\tau_{14})G(\tau_{32})G(\tau_{34})^{\frac{q}{2}-2}G(\tau_{43})^{\frac{q}{2}}\Big )\ . \label{kernel}
\end{aligned}
\end{equation}
By summing the geometric series, the full four-point function reduces to  
\begin{align}
\mathcal{F}(\tau_1,\tau_2; \tau_3, \tau_4) = \int_{0}^{\beta} d\tau_a d\tau_b [\mathbf{1}-K]^{-1}(\tau_1,\tau_2;\tau_a,\tau_b) \mathcal{F}_{0}(\tau_a, \tau_b; \tau_3, \tau_4)\,. \label{4pt1}
\end{align}

In the case of zero chemical potential $\mu=0$ the cSYK four-point function is the sum of anti-symmetric and symmetric parts and was computed in the conformal limit in \cite{Bulycheva:2017uqj, Peng:2017spg} (see also \cite{Murugan:2017eto}). 
In this case the anti-symmetric part of the cSYK four-point function coincides with the four-point function of the Majorana SYK, which was initially computed in \cite{Maldacena:2016hyu, Polchinski:2016xgd}.
To proceed we define anti-symmetric and symmetric left and right eigenfunctions of the kernel as $\Phi^{a\textrm{L}}_{h,m}(\tau_{1},\tau_{2}), \Phi^{s\textrm{L}}_{h,m}(\tau_{1},\tau_{2})$ and $\Phi^{a\textrm{R}}_{h,m}(\tau_{1},\tau_{2}), \Phi^{s\textrm{R}}_{h,m}(\tau_{1},\tau_{2})$, so 
\begin{equation}
\begin{aligned}
&\int_{0}^{\beta} d\tau_{3}d\tau_{4}\, K(\tau_{1},\tau_{2};\tau_{3},\tau_{4}) \Phi^{\sigma,\textrm{R}}_{h,m}(\tau_{3},\tau_{4}) = k^{\sigma}_{h,m}\Phi^{\sigma\textrm{R}}_{h,m}(\tau_{1},\tau_{2})\,, \\
&\int_{0}^{\beta} d\tau_{1}d\tau_{2} \, \Phi^{\sigma\textrm{L}*}_{h,m}(\tau_{1},\tau_{2}) K(\tau_{1},\tau_{2};\tau_{3},\tau_{4})  = k^{\sigma}_{h,m}\Phi^{\sigma\textrm{L}*}_{h,m}(\tau_{3},\tau_{4})\,,
\end{aligned}
\end{equation}
where $\sigma=a, s$ and $k^{\sigma}_{h,m}$ are the eigenvalues. The eigenfunctions are normalized as 
\begin{align}\label{eq:NormDef}
\int_{0}^{\beta} d\tau_{1} d\tau_{2} \Phi_{h,m}^{\sigma\textrm{L}*}(\tau_{1},\tau_{2})\Phi_{h',m'}^{\sigma'\textrm{R}}(\tau_{1},\tau_{2}) = \mathcal{N}_{h,m}^{\sigma}\delta^{\sigma \sigma'}\delta_{hh'}\delta_{mm'}\,,
\end{align}
where $\mathcal{N}_{h,m}^{\sigma}$ are  normalization constants.
The kernel is translation invariant, and therefore its eigenfunctions carry an integer quantum number $m$ and satisfy $\Phi_{h,m}^{\sigma\textrm{R}}(\tau_{1},\tau_{2}) = e^{im(\tau_{1}+\tau_{2})/\beta} \phi^{\sigma\textrm{R}}_{h,m}(\tau_{12})$. The index $h$ labels the eigenfunctions and eigenvalues within a given $m$ sector, and in general its allowed range depends on $m$.

In the conformal limit, we replace the exact Green’s function $G(\tau)$ in the kernel $K$ in (\ref{kernel}) by its conformal approximation $G_c(\tau)$ in (\ref{GcSol}), thereby obtaining the conformal kernel $K_c$. In this limit, $h(h-1)$ is the eigenvalue of the $SL(2)$ Casimir operator \cite{Maldacena:2016hyu,Kitaev:2017awl, Bulycheva:2017uqj, Peng:2017spg}, and $h$ takes positive even values $h=2,4,6,8,\dots$ for $\sigma=a$ and positive odd values $h=1,3,5,7,\dots$ for $\sigma=s$, as well as continuous values (the principal series) $h=1/2+is$ with $s>0$ for both $\sigma=a,s$. For the discrete series, $m$ must satisfy $|m|\geq h$, and therefore in the $m=0$ sector only the principal series $h=1/2+is$, with $s>0$, is allowed.

For zero chemical potential, $\mu=0$, the particle-hole symmetry of the Green’s function, $G(\tau)=G(\beta-\tau)$, allows one to symmetrize the kernel $K$ \cite{Maldacena:2016hyu, Gu:2019jub}. Since $G(\tau)$ is real, the kernel is a real, symmetric operator, and thus its eigenvalues $k^{\sigma}_{h,m}$ are real. For non-zero chemical potential, the kernel is still real since $G(\tau)$ is real, but it cannot be symmetrized while keeping it real.
As a result, its eigenvalues are either real or come in complex-conjugate pairs. This is a simple case of $\mathcal{PT}$ symmetry, where $\mathcal{P}$ is the identity, and the time-reversal operator $\mathcal{T}$ acts as $i\to -i$ \cite{Bender:2023cem}.  Since the eigenvalues of the kernel are real at $\mu=0$, they can become complex as $\mu$ is increased only if they collide with one another. This phenomenon occurs, for example, in the well-known flow from the quantum Ising model to the Yang–Lee model \cite{vonGehlen:1991zlm, ArguelloCruz:2025zuq}.

The kernel eigenfunctions $\Phi_{h,m}^{\sigma \textrm{R}}$ and $\Phi_{h,m}^{\sigma \textrm{L}}$ for $\mu=0$ form a complete orthogonal basis of bilocal functions and therefore satisfy the completeness (closure) relation.
\begin{align}
\sum_{\sigma=a,s}\sum_{h,m} \frac{1}{\mathcal{N}_{h,m}^{\sigma}}\Phi_{h,m}^{\sigma \textrm{R}}(\tau_{1},\tau_{2})\Phi_{h,m}^{\sigma\textrm{L}*}(\tau_{3},\tau_{4}) = \delta(\tau_{13})\delta(\tau_{24})\,. \label{CompRel1}
\end{align}
We can use the completeness relation (\ref{CompRel1}) in \eqref{4pt1} for a general case of non-zero $\mu$ to find
\begin{equation}
\begin{aligned}
\mathcal{F}(\tau_{1},\tau_{2};\tau_{3},\tau_{4})=&\sum_{\sigma=a,s}\sum_{h,m}\int_{0}^{\beta} d\tau_{a} d\tau_{b} d\tau_{c} d\tau_{d} \, [\textbf{1}-K]^{-1} (\tau_1, \tau_2; \tau_{a}, \tau_{b}) \times \\
& \qquad\qquad\quad \times\frac{1}{\mathcal{N}_{h,m}^{\sigma}}\Phi_{h,m}^{\sigma\textrm{R}}(\tau_{a},\tau_{b})\Phi_{h,m}^{\sigma\textrm{L}*}(\tau_{c},\tau_{d}) \mathcal{F}_{0}(\tau_{c},\tau_{d};\tau_{3},\tau_{4})\,. \label{4ptexact1}
\end{aligned}
\end{equation}
For non-zero $\mu$ the functions $\Phi_{h,m}^{\sigma\textrm{R}}$ and $\Phi_{h,m}^{\sigma \textrm{L}}$ are no longer eigenfunctions of $K$, and are mixed under the action of the kernel.  However, it will be convenient to still use this basis.

\section{Four-point function in the conformal limit}
\label{SecConfLim}

Below, we compute the four-point function in the conformal limit at non-zero $\mu$, using (\ref{4ptexact1}), where we replace the exact large-$N$ kernel $K$ and the eigenfunctions at $\mu=0$ with the conformal kernel $K_c$ and the conformal eigenfunctions. We denote  this four-point function by $\mathcal{F}_{c}$. We will see later that the result agrees with the computation based on the CFT analysis, and it will allow us to extract the structure constants involving two complex fermions and the bilocal operators of the cSYK model for arbitrary $\mathscr{E}$.

At non-zero $\mathscr{E}$ we introduce a set of right functions $W^{\sigma}_{h}$ with $\sigma=a,s$ at finite temperature $T=1/\beta$ as 
\begin{equation}
\begin{aligned}
W_{h}^{a}(\tau_{1},\tau_{2};\tau_{0}) = \frac{\sgn(\tau_{12})
e^{2\pi \mathscr{E}\left(\frac{|\tau_{12}|}{\beta}-\frac{1}{2}\right)\sgn(\tau_{12})} }{|\frac{\beta}{\pi}\sin(\frac{\pi \tau_{12}}{\beta})|^{2\Delta-h}|\frac{\beta}{\pi}\sin(\frac{\pi \tau_{10}}{\beta})|^{h}|\frac{\beta}{\pi}\sin(\frac{\pi \tau_{20}}{\beta})|^{h}}\,,  \\
W_{h}^{s}(\tau_{1},\tau_{2};\tau_{0}) = \frac{\sgn(\tau_{10})\sgn(\tau_{20})
e^{2\pi \mathscr{E}\left(\frac{|\tau_{12}|}{\beta}-\frac{1}{2}\right)\sgn(\tau_{12})} }{|\frac{\beta}{\pi}\sin(\frac{\pi \tau_{12}}{\beta})|^{2\Delta-h}|\frac{\beta}{\pi}\sin(\frac{\pi \tau_{10}}{\beta})|^{h}|\frac{\beta}{\pi}\sin(\frac{\pi \tau_{20}}{\beta})|^{h}} \,,\label{WhaWhs}
\end{aligned}
\end{equation}
and a set of left functions  $\tilde{W}^{\sigma}_{h}$ as 
\begin{equation}
\begin{aligned}
\tilde{W}_{h}^{a}(\tau_{1},\tau_{2};\tau_{0}) = \frac{\sgn(\tau_{12})
e^{-2\pi \mathscr{E}\left(\frac{|\tau_{12}|}{\beta}-\frac{1}{2}\right)\sgn(\tau_{12})} }{|\frac{\beta}{\pi}\sin(\frac{\pi \tau_{12}}{\beta})|^{2-2\Delta-h}|\frac{\beta}{\pi}\sin(\frac{\pi \tau_{10}}{\beta})|^{h}|\frac{\beta}{\pi}\sin(\frac{\pi \tau_{20}}{\beta})|^{h}}\,,  \\
\tilde{W}_{h}^{s}(\tau_{1},\tau_{2};\tau_{0}) = \frac{\sgn(\tau_{10})\sgn(\tau_{20})
e^{-2\pi \mathscr{E}\left(\frac{|\tau_{12}|}{\beta}-\frac{1}{2}\right)\sgn(\tau_{12})} }{|\frac{\beta}{\pi}\sin(\frac{\pi \tau_{12}}{\beta})|^{2-2\Delta-h}|\frac{\beta}{\pi}\sin(\frac{\pi \tau_{10}}{\beta})|^{h}|\frac{\beta}{\pi}\sin(\frac{\pi \tau_{20}}{\beta})|^{h}} \,.\label{WthaWths}
\end{aligned}
\end{equation}
At $\mathscr{E}=0$ ($\mu=0$) we can obtain the eigenfunctions  $\Phi^{\sigma R}_{h,m}$ and $\Phi^{\sigma L}_{h,m}$ of the conformal kernel $K_{c}$ as \cite{Kitaev:2017awl}
\begin{align}
\Phi_{h,m}^{\sigma R} = \int_{0}^{\beta}\frac{d\tau_{0}}{\beta} W_{h}^{\sigma}(\tau_{1},\tau_{2};\tau_{0})e^{i\frac{2\pi m}{\beta}\tau_{0}}, \quad
\Phi_{h,m}^{\sigma L*} = \int_{0}^{\beta}\frac{d\tau_{0}}{\beta} \tilde{W}_{1-h}^{\sigma}(\tau_{1},\tau_{2};\tau_{0})e^{-i\frac{2\pi m}{\beta}\tau_{0}}\,. \label{PhiasE0}
\end{align}
It is known that these eigenfunctions are normalizable for $\sigma  = a$ only when $h=2,4,6,\dots$ and $h=1/2+is$, $s>0$ and  for $\sigma = s$ only when $h=1,3,5,\dots$ and $h=1/2+is, s>0$ \cite{Maldacena:2016hyu, Kitaev:2017hnr, Kitaev:2017awl} and their inner product is 
\begin{align}
\int_{0}^{\beta} d\tau_{1}d\tau_{2}\,\Phi_{h,m}^{\sigma L*}(\tau_{1},\tau_{2})
\Phi_{h',m'}^{\sigma' R}(\tau_{1},\tau_{2}) = \frac{1}{\beta}\mathcal{N}_{h} \delta^{\sigma\sigma'} \delta_{hh'}\delta_{mm'}\,,
\end{align}
where the normalization constant $\mathcal{N}_{h}$ does not depend on $\sigma$ nor $m$, and reads
\begin{align}
 \mathcal{N}_{h}= \begin{cases}
\frac{2\pi^{2}}{h-1/2}, \qquad\; h \in \mathbbm{Z}_{+} \\
\frac{2\pi \tan(\pi h)}{h-1/2}, \quad h = 1/2+is,\; s>0
\end{cases}\,, \label{NormConst}
\end{align}
where for the principal series $h=1/2+is$ we assume that $\delta_{hh'}=2\pi\delta(s-s')$.
The eigenvalues of the conformal kernel do not depend on $m$ and at $\mathscr{E}=0$ ($\mu= 0$) we find
\begin{align}
\int_{0}^{\beta} d\tau_{3}d\tau_{4}K_{c}(\tau_{1},\tau_{2};\tau_{3},\tau_{4})\Phi_{h,m}^{\sigma R}(\tau_{3},\tau_{4}) = k^{\sigma}_{c}(h)\Phi_{h,m}^{\sigma R}(\tau_{1},\tau_{2})\,,
\end{align}
where $k_{c}^{a}(h)$ and $k_{c}^{s}(h)$ are \cite{Maldacena:2016hyu, Kitaev:2017awl, Bulycheva:2017uqj, Gu:2019jub}
\begin{equation}
\begin{aligned}
&k_{c}^{a}(h) = \frac{\Gamma(2\Delta- h)\Gamma(2\Delta +h-1)}{\Gamma(2\Delta-2)\Gamma(2\Delta+1)}\Big(1-\frac{\sin(\pi h)}{\sin(2\pi \Delta)}\Big),  \\ 
&k_{c}^{s}(h) = \frac{\Gamma(2\Delta- h)\Gamma(2\Delta +h-1)}{\Gamma(2\Delta-1)\Gamma(2\Delta)}\Big(1+\frac{\sin(\pi h)}{\sin(2\pi \Delta)}\Big)\,. \label{kacksc}
\end{aligned}
\end{equation}

Using the representation (\ref{PhiasE0}) for the eigenfunctions at $\mathscr{E}=0$ in the completeness relation (\ref{CompRel1}), and taking the sum over the index $m$, we find \cite{Kitaev:2017awl}
\begin{equation}
\begin{aligned}
&\int_{0}^{\beta} d\tau_{0}  \bigg(\sum_{h=h^{a}}\frac{1}{\mathcal{N}_{h}}W_{h}^{a}(\tau_{1},\tau_{2};\tau_{0})\tilde{W}_{1-h}^{a}(\tau_{3},\tau_{4};\tau_{0})  \\
& \qquad\qquad \qquad +\sum_{h=h^{s}}\frac{1}{\mathcal{N}_{h}}W_{h}^{s}(\tau_{1},\tau_{2};\tau_{0})\tilde{W}_{1-h}^{s}(\tau_{3},\tau_{4};\tau_{0}) \bigg)  = \delta(\tau_{13})\delta(\tau_{24})\,, \label{CompRel}
\end{aligned}
\end{equation}
where we introduced short notations for the sums over $h$:
\begin{align}
\sum_{h=h^{a}} \equiv \int_{1/2+i0}^{1/2+i\infty}\frac{dh}{2\pi i} +  \sum_{h=2,4,6,\dots} , \quad 
\sum_{h=h^{s}} \equiv \int_{1/2+i0}^{1/2+i\infty}\frac{dh}{2\pi i} + \sum_{h=1,3,5,\dots} \,. \label{sumsofh}
\end{align}

We now note that, for a non-zero asymmetry parameter $\mathscr{E}$, the completeness relation (\ref{CompRel}) still holds, since the factors $\exp(\pm 2\pi \mathscr{E}(\dots))$ in (\ref{WhaWhs}) and (\ref{WthaWths}) multiply the delta functions and cancel each other. Therefore, in the conformal limit we can compute the four-point function at non-zero $\mathscr{E}$ using (\ref{4ptexact1}) together with the completeness relation (\ref{CompRel}), with $W_h^{\sigma}$ and $\tilde{W}_h^{\sigma}$ given in (\ref{WhaWhs}) and (\ref{WthaWths}):
\begin{equation}
\begin{aligned}
\mathcal{F}_{c}(\tau_{1},\tau_{2};\tau_{3},\tau_{4})&=\int_{0}^{\beta} d\tau_{0} \int_{0}^{\beta} d\tau_{a} d\tau_{b} d\tau_{c} d\tau_{d} \, [\textbf{1}-K_{c}]^{-1} (\tau_1, \tau_2; \tau_{a}, \tau_{b}) \times  \\
& \times\bigg(\sum_{h=h^{a}}\frac{1}{\mathcal{N}_{h}}W_{h}^{a}(\tau_{a},\tau_{b};\tau_{0})\tilde{W}_{1-h}^{a}(\tau_{c},\tau_{d};\tau_{0})  \\
 &\quad+\sum_{h=h^{s}}\frac{1}{\mathcal{N}_{h}}W_{h}^{s}(\tau_{a},\tau_{b};\tau_{0})\tilde{W}_{1-h}^{s}(\tau_{c},\tau_{d};\tau_{0}) \bigg) \mathcal{F}_{0}(\tau_{c},\tau_{d};\tau_{3},\tau_{4})\,. \label{4ptexact2}
\end{aligned}
\end{equation}
In order to proceed with the computation, we  find the action of the conformal kernel $K_{c}$ on the right functions $W_{h}^{a}$ and $W_{h}^{s}$  as well as the action of the bare four-point function $\mathcal{F}_{0}$ on the left functions $\tilde{W}_{1-h}^{a}$ and $\tilde{W}_{1-h}^{s}$. The explicit computations can be found in Appendix \ref{AppDiagcSYK}.
At non-zero $\mathscr{E}$ these actions mix the $a$ and $s$ sectors  and we can represent them in matrix form as 
\begin{equation}
\begin{aligned}
  \left(  \begin{array}{cc}
    K_{c} *W_{h}^{a} & K_{c} *W_{h}^{s} \\ 
  \end{array}\right) &= \left(  \begin{array}{cc}
    W_{h}^{a}(\tau_{1},\tau_{2};\tau_{0}) & W_{h}^{s}(\tau_{1},\tau_{2};\tau_{0}) \\ 
  \end{array}\right) K_{c}(h)\,, \quad \\
\left(  \begin{array}{c}
    \tilde{W}_{1-h}^{a}*\mathcal{F}_{0} \\ \tilde{W}_{1-h}^{s}*\mathcal{F}_{0} \\ 
  \end{array}\right) &= \frac{1}{b}\Big(\frac{b}{J^2}\Big)^{2\Delta} K_{c}(h)\Lambda \left(  \begin{array}{c}
    W_{1-h}^{a}(\tau_{3},\tau_{4};\tau_{0})  \\ 
    W_{1-h}^{s}(\tau_{3},\tau_{4};\tau_{0}) \\ 
  \end{array}\right)\,,
\end{aligned}
\end{equation}
where the $2\times 2$ matrices $K_{c}(h)$ and $\Lambda$ are 
\begin{align}
K_{c}(h) =  \left(  \begin{array}{cc}
    K_{h}^{aa} & K_{h}^{as} \\ 
    K_{h}^{sa} & K_{h}^{ss} \\ 
  \end{array}\right)\,, \quad 
  \Lambda =  \left(  \begin{array}{cc}
    \frac{1}{q-1} & 0 \\ 
    0 & -1 \\ 
  \end{array}\right)\,, \label{KcandLam}
 \end{align}
and we defined $K_{c}*W_{h}^{\sigma} \equiv \int d\tau_{3}d\tau_{4}K_{c}(\tau_{1},\tau_{2};\tau_{3},\tau_{4})W_{h}^{\sigma}(\tau_{3},\tau_{4};\tau_{0})$ and also $\tilde{W}_{1-h}^{\sigma}*\mathcal{F}_{0} \equiv \int d\tau_{1}d\tau_{2}\tilde{W}_{1-h}^{\sigma}(\tau_{1},\tau_{2};\tau_{0})\mathcal{F}_{0}(\tau_{1},\tau_{2};\tau_{3},\tau_{4})$
and the coefficients $K_{h}^{\sigma \sigma'}$ with $\sigma,\sigma' = a,s$ are computed in (\ref{KssprRes}).

Now, using $K_c * W_h^{\sigma} = W_h^{\sigma'}(\tau_{1},\tau_{2};\tau_{0}) K_h^{\sigma'\sigma}$, with summation over $\sigma'$ implied, we find
\begin{align}
[\textbf{1}-K_{c}]^{-1}*W^{\sigma}_{h}  = W^{\sigma'}_{h}(\tau_{1},\tau_{2};\tau_{0}) ([\textbf{1}-K_{c}(h)]^{-1})^{\sigma' \sigma}\,,
\end{align}
and thus, we can write the conformal four-point function in (\ref{4ptexact2})
as the sum of four terms 
\begin{align}
\mathcal{F}_{c}(\tau_{1},\tau_{2};\tau_{3},\tau_{4}) = \sum_{\sigma,\sigma'=a,s}
\mathcal{F}_{c}^{\sigma \sigma'}(\tau_{1},\tau_{2};\tau_{3},\tau_{4})\,, \label{4ptAsSum}
\end{align}
where each term can be expressed as 
\begin{equation}
\begin{aligned}
\mathcal{F}^{\sigma \sigma'}_{c}(\tau_{1},\tau_{2};\tau_{3},\tau_{4})=& \frac{1}{b}\Big(\frac{b}{J^2}\Big)^{2\Delta}\bigg(\sum_{h=h^{a}}  \frac{1}{\mathcal{N}_{h}}([\textbf{1}-K_{c}(h)]^{-1})^{\sigma a} (K_{c}(h)\Lambda)^{a\sigma'} \\
 &+\sum_{h=h^{s}} \frac{1}{\mathcal{N}_{h}}([\textbf{1}-K_{c}(h)]^{-1})^{\sigma s} (K_{c}(h)\Lambda)^{s\sigma'}   \bigg)I_{h}^{\sigma \sigma'}(\tau_{1},\tau_{2};\tau_{3},\tau_{4}) \,, \label{4ptexact3}
\end{aligned}
\end{equation}
and the integrals $I_{h}^{\sigma\sigma'}$ with $\sigma,\sigma' = a,s$ are defined by
\begin{align}
I_{h}^{\sigma\sigma'}(\tau_{1},\tau_{2};\tau_{3},\tau_{4}) \equiv  \int_{0}^{\beta} d\tau_{0} W_{h}^{\sigma}(\tau_{1},\tau_{2}; \tau_{0}) W_{1-h}^{\sigma'}(\tau_{3},\tau_{4};\tau_{0})\,. \label{IntegralsDef0}
\end{align}
We compute these integrals in Appendix \ref{IntegralsAS} and the final result can be written as 
\begin{equation}
\begin{aligned}
\frac{\mathcal{F}^{\sigma \sigma'}_{c}(\tau_{1},\tau_{2};\tau_{3},\tau_{4})}{G_{c}(\tau_{12})G_{c}(\tau_{34})}= &\alpha_{0}^{\sigma \sigma'}\bigg(\sum_{h = h^{a}} \frac{1}{\mathcal{N}_{h}}([\textbf{1}-K_{c}(h)]^{-1})^{\sigma a} (K_{c}(h))^{a\sigma'} \\
&\quad +\sum_{h=h^{s}}\frac{1}{\mathcal{N}_{h}}([\textbf{1}-K_{c}(h)]^{-1})^{\sigma s} (K_{c}(h))^{s\sigma'}   \bigg)  \Psi_{h}^{\sigma \sigma'}(\chi) \,, \label{4ptFinal1}
\end{aligned}
\end{equation}
where we introduced the factors $\alpha_{0}^{\sigma \sigma'}$, defined by
\begin{align}
\alpha_{0}^{aa}= \frac{1}{b(q-1)}, \quad \alpha_{0}^{as}= -\frac{\sgn(\tau_{13}\tau_{14}\tau_{34})}{b}, \quad 
\alpha_{0}^{sa}=\frac{\sgn (\tau_{12}\tau_{13}\tau_{23})}{b(q-1)},\quad \alpha_{0}^{ss}=-\frac{1}{b}\,,
\end{align}
and the explicit expressions for $\Psi_{h}^{\sigma \sigma'}(\chi)$ with $\sigma,\sigma' =a,s$ are 
\begin{equation}
\begin{aligned}
    \Psi_{h}^{aa}(\chi)&=\left\lbrace
    \begin{matrix}
       \frac{\tan(\pi h)}{\tan (\frac{\pi h}{2})}\frac{\Gamma(h)^2}{\Gamma(2h)}\chi^h\,_2F_1(h,h;2h;\chi)+(h\to 1-h) &,&0<\chi<1\\
       \frac{2\Gamma\left(\frac{h}{2}\right)\Gamma\left(\frac{1-h}{2}\right)}{\sqrt{\pi}}\,_2F_1\left(\frac{h}{2},\frac{1-h}{2};\frac{1}{2};\frac{(2-\chi)^2}{\chi^2}\right) &,&1<\chi<2
    \end{matrix}
    \right.\\
\Psi_{h}^{as}(\chi)&=\left\lbrace
    \begin{matrix}
       \frac{\tan(\pi h)}{\cot (\frac{\pi h}{2})}\left[\frac{\Gamma(h)^2}{\Gamma(2h)}\chi^h\,_2F_1(h,h;2h;\chi) - (h \to 1-h) \right]&,&0<\chi<1\\
       0 &,&1<\chi<2
    \end{matrix}
    \right.\\
\Psi_{h}^{sa}(\chi)&=\left\lbrace
    \begin{matrix}
       \frac{\tan(\pi h)}{\tan (\frac{\pi h}{2})} \left[\frac{\Gamma(h)^2}{\Gamma(2h)}\chi^h\,_2F_1(h,h;2h;\chi)- (h \to1-h) \right]&,&0<\chi<1\\
       0 &,&1<\chi<2
    \end{matrix}
    \right.\\
\Psi_{h}^{ss}(\chi)&=\left\lbrace
    \begin{matrix}
       \frac{\tan(\pi h)}{\cot (\frac{\pi h}{2})}\frac{\Gamma(h)^2}{\Gamma(2h)}\chi^h\,_2F_1(h,h;2h;\chi)+( h \to 1-h) &,&0<\chi<1\\
      -\frac{4\Gamma\left(1-\frac{h}{2}\right)\Gamma\left(\frac{h+1}{2}\right)}{\sqrt{\pi}} \left(\frac{2-\chi}{\chi}\right)\,_2F_1\left(1-\frac{h}{2},\frac{h+1}{2};\frac{3}{2};\frac{(2-\chi)^2}{\chi^2}\right) &,&1<\chi<2
    \end{matrix}
    \right., \label{PsiaaToPsiss}
\end{aligned}
\end{equation}
where $_{2}F_{1}(a,b;c;z)$ is the hypergeometric function and  the cross-ratio $\chi$ at finite temperature is 
\begin{align}
\chi = \frac{\sin(\frac{\pi \tau_{12}}{\beta})\sin(\frac{\pi \tau_{34}}{\beta})}{\sin(\frac{\pi \tau_{13}}{\beta})\sin(\frac{\pi \tau_{24}}{\beta})}\,.
\end{align}
Below, we consider only the domain $0<\chi<2$, since values of $\chi$ outside this range can be obtained through the map $\chi \to \chi/(\chi-1)$, as discussed in Appendix~\ref{IntegralsAS}.

For $\mathscr{E}=0$ we have $K_{c}(h)=\textrm{diag}(k_{c}^{a}(h), k_{c}^{s}(h))$, where $k_{c}^{\sigma}(h)$ are defined in (\ref{kacksc}). Since in this case  $K_{c}(h)$ is a diagonal matrix, we see from (\ref{4ptFinal1}) that  $\mathcal{F}^{as}=\mathcal{F}^{sa}=0$ and only the terms $\mathcal{F}^{aa}$ and $\mathcal{F}^{ss}$ survive. Thus we have for the four-point function
\begin{align}
\frac{\mathcal{F}_{c}(\tau_{1},\tau_{2};\tau_{3},\tau_{4})}{G_{c}(\tau_{12})G_{c}(\tau_{34})}= \alpha_{0}^{aa}\sum_{h=h^{a}}\frac{1}{\mathcal{N}_{h}}\frac{k_{c}^{a}(h)}{1-k_{c}^{a}(h)} \Psi_{h}^{aa}(\chi)+\alpha_{0}^{ss}\sum_{h=h^{s}}\frac{1}{\mathcal{N}_{h}}\frac{k_{c}^{s}(h)}{1-k_{c}^{s}(h)} \Psi_{h}^{ss}(\chi)\,. \label{4ptE0}
\end{align}
The term $\mathcal{F}^{aa}$ of the four-point function (\ref{4ptE0}) agrees with the results in \cite{Maldacena:2016hyu} (eq. (3.83)) and \cite{Bulycheva:2017uqj} (eq. (5.43)) and the term $\mathcal{F}^{ss}$ agrees with the result in \cite{Bulycheva:2017uqj} (eq. (5.44)) up to a factor of $(q-1)$
\footnote{We note that  $\Psi_{h}^{aa}(\chi)$ in (\ref{PsiaaToPsiss}) is related to $\Psi_{h}(\chi)$ in \cite{Maldacena:2016hyu} (eq. (3.74))  as 
$\Psi_{h}^{aa}(\chi) = 2 \Psi_{h}(\chi)$, and  coincides with $\Psi_{h}^{A}(\chi)$  defined in \cite{Bulycheva:2017uqj} (eqs. (5.15)) and $\Psi_{h}^{ss}(\chi)$ coincides with $\Psi_{h}^{S}(\chi)$  defined in \cite{Bulycheva:2017uqj} (eqs. (5.16)). We also note that in \cite{Maldacena:2016hyu} one has $\left\langle 4\textrm{pt} \right\rangle_{\textrm{SYK}} = G_{12}G_{34} + \mathcal{F}_{\textrm{SYK}}/N_{\textrm{SYK}} +\dots$, where $N_{\textrm{SYK}}$ is the number of Majorana fermions, whereas in our case  $\left\langle 4\textrm{pt} \right\rangle_{\textrm{cSYK}} = G_{12}G_{34} + \mathcal{F}_{\textrm{cSYK}}/N_{\textrm{cSYK}}+\dots$, where $N_{\textrm{cSYK}}$ is the number of complex fermions. Since $N_{\textrm{SYK}}=2N_{\textrm{cSYK}}$ the four-point functions are related as $\mathcal{F}_{\textrm{cSYK}}^{aa} =\mathcal{F}_{\textrm{SYK}}/2$. A similar factor of $2$ discrepancy appears in   \cite{Bulycheva:2017uqj} since $\mathcal{F}_{0}^{A/S}$ in their eqs. (5.38), (5.39) miss a factor of $1/2$ and thus $\mathcal{F}_{0}^{A}-\mathcal{F}_{0}^{S}$ is twice as large as our  $\mathcal{F}_{0}$, therefore $\mathcal{F}_{\textrm{cSYK}}^{aa} =\mathcal{F}^{A}/2$. For the term $\mathcal{F}_{\textrm{cSYK}}^{ss}$ we find $\mathcal{F}_{\textrm{cSYK}}^{ss} =(q-1)\mathcal{F}^{S}/2$, where $\mathcal{F}^{S}$ is given in \cite{Bulycheva:2017uqj} (eqs. (5.44)). }.  In the next section we show that the four-point function (\ref{4ptAsSum}), (\ref{4ptFinal1}), can be written in the form which agrees with the result obtained using CFT analysis, providing an additional consistency check.

\section{Four-point function as a sum over conformal blocks}
\label{4ptAsBlocks}
To proceed, we first discuss properties of the matrix $K_{c}(h)$ defined in (\ref{KcandLam}), whose entries $K_{h}^{\sigma\sigma'} \equiv (K_{c}(h))^{\sigma\sigma'}$, with $\sigma,\sigma' = a,s$, are given in (\ref{KssprRes}). Since the matrix $K_{c}(h)$ is not symmetric, its left and right eigenvectors are not the same. Let us denote the right eigenvectors of $K_{c}(h)$ by $\mathbf{v}_{1}=(v_{1}^{a},v_{1}^{s})$ and $\mathbf{v}_{2}=(v_{2}^{a},v_{2}^{s})$, and its left eigenvectors by $\mathbf{w}_{1}=(w_{1}^{a},w_{1}^{s})$ and $\mathbf{w}_{2}=(w_{2}^{a},w_{2}^{s})$, so 
\begin{align}
K_{c}(h)\textbf{v}_{i}=k_{i}(h)\textbf{v}_{i}, \quad \textbf{w}^{T}_{i}K_{c}(h) =\textbf{w}^{T}_{i} k_{i}(h), \quad i = 1,2\,,
\end{align}
where the eigenvalues $k_{1}(h)$ and $k_{2}(h)$ are given by the formula 
\begin{equation}
\begin{aligned}
k_{1,2}(h) &= \frac{\Gamma(2\Delta-h)\Gamma(2\Delta+h-1)}{\Gamma(2\Delta+1)\Gamma(2\Delta-1)}\left(2\Delta-1+\frac{\cos(2\theta)\sin(\pi h)}{\sin(2\pi \Delta)}\mp \sqrt{P}\right)\,, \\
P&\equiv  \sin^{2}(2\theta)\left(1-\frac{\sin^{2}(\pi h)}{\sin^{2}(2\pi\Delta)}\right)+
\left(\cos(2\theta)+(2\Delta-1)\frac{\sin(\pi h)}{\sin(2\pi \Delta)}\right)^{2}\,, \label{k1k2eigs}
\end{aligned}
\end{equation}
and here we used the asymmetry parameter $\theta$ instead of $\mathscr{E}$. The right eigenvectors $\textbf{v}_{1,2}$ read
\begin{align}
\textbf{v}_{1,2}(h,\theta) = \frac{1/\sqrt{2}}{1+(2\Delta-1)\frac{\sin(\pi h)}{\sin(2\pi \Delta)}} S\left(  \begin{array}{c}
   \frac{\sin(2\theta)}{\sin(2\pi \Delta)}(2\Delta-1+\cos(\pi h)) \pm \sqrt{P}   \\ 
   1+\frac{\sin(2\theta)}{\sin(2\pi \Delta)} + (2\Delta-1)\frac{\sin(\pi h+ 2\theta)}{\sin(2\pi \Delta)}  \\ 
  \end{array}\right)\,, \label{v1v2right}
\end{align}
where the orthogonal matrix $S$ is given by\footnote{We note that the matrix $K_{c}(h)$ is related to the matrix $K_{G}(h)$ in \cite{Tikhanovskaya:2020elb}  as $K_{c}(h)=SK_{G}(1-h)S^{T}$, therefore the eigenvalues $k_{1,2}(h)$ coincide with those in \cite{Tikhanovskaya:2020elb} (eq. (4.25)).}  
\begin{align}
S= \frac{1}{\sqrt{2}}  
\left(\begin{array}{cc}
   1 & 1 \\ 
   -1  &1 \\ 
  \end{array}\right)\,. \label{Smat}
\end{align}
We can write the left eigenvectors $\textbf{w}_{1}$ and $\textbf{w}_{2}$ in the form 
\begin{align}
\textbf{w}_{1}^{T}(h,\theta) = \frac{\textbf{v}_{2}^{T}(h,-\theta)\sigma_{x}}{\textbf{v}_{2}^{T}(h,-\theta)\sigma_{x} \textbf{v}_{1}(h,\theta)}, \quad 
\textbf{w}_{2}^{T}(h,\theta) = \frac{\textbf{v}_{1}^{T}(h,-\theta)\sigma_{x}}{\textbf{v}_{1}^{T}(h,-\theta)\sigma_{x} \textbf{v}_{2}(h,\theta)}\,, \label{w1w2left}
\end{align}
where $\sigma_{x}$ is the Pauli $x$ matrix.  One can check that the left and right eigenvectors are orthonormal
\begin{align}
\textbf{w}_{i}^{T}(h,\theta) \textbf{v}_{j}(h,\theta)= \delta_{ij}, \quad i,j=1,2\,.
\end{align}
 At $\theta =0$ ($\mathscr{E}=0$), we find $k_{1}(h)=k^{a}_{c}(h)$ and $k_{2}(h)=k^{s}_{c}(h)$, where 
$k^{a}_{c}(h)$ and $k^{s}_{c}(h)$ are written in (\ref{kacksc}) and we chose normalization in (\ref{v1v2right}) such that 
\begin{align}
\textbf{v}_{1}(h,0)=\left(  \begin{array}{c}
   1  \\ 
   0 \\ 
  \end{array}\right), \;\;  \textbf{v}_{2}(h,0)=\left(  \begin{array}{c}
   0  \\ 
   1 \\ 
  \end{array}\right), \quad 
  \textbf{w}_{1}^{T}(h,0)=\left(  \begin{array}{cc}
   1 & 0  \\ 
  \end{array}\right), \;\;  \textbf{w}^{T}_{2}(h,0)=\left(  \begin{array}{cc}
   0  & 1 \\    
  \end{array}\right)\,.
\end{align}
 
 We note the relation 
\begin{align}
K_{c}(1-h) = \Lambda K_{c}(h)^{T} \Lambda^{-1}\,, \label{RelKc1}
\end{align}
where the matrix $\Lambda$ is defined in (\ref{KcandLam}), and therefore the eigenvalues satisfy $k_{1,2}(1-h)=k_{1,2}(h)$. Using the left and right eigenvectors, we can write $K_{h}^{\sigma \sigma'}=k_{1}(h)v_{1}^{\sigma}w_{1}^{\sigma'} +k_{2}(h)v_{2}^{\sigma}w_{2}^{\sigma'}$  for $\sigma,\sigma'=a,s$ and similarly 
\begin{align}
([\textbf{1}-K_{c}(h)]^{-1})^{\sigma \sigma'} = \frac{v_{1}^{\sigma}w_{1}^{\sigma'}}{1-k_{1}(h)}+ \frac{v_{2}^{\sigma}w_{2}^{\sigma'}}{1-k_{2}(h)}\,. 
\end{align}
Hence, for integer values of $h$ we find
\begin{equation}
\begin{aligned}
&([\textbf{1}-K_{c}(h)]^{-1})^{aa} = \frac{1}{1-k_{1}(h)}, \;\;   ([\textbf{1}-K_{c}(h)]^{-1})^{sa} = 0, \quad h = 2,4,6,\dots \,, \\
&([\textbf{1}-K_{c}(h)]^{-1})^{ss} = \frac{1}{1-k_{2}(h)}, \;\;    ([\textbf{1}-K_{c}(h)]^{-1})^{as} = 0, \quad h = 1,3,5,\dots \,, \label{KcDiscr}
\end{aligned}
\end{equation}
where we used the results for $\textbf{v}_{1,2}$, $\textbf{w}_{1,2}$ in Appendix $\ref{AppKc}$. 
Moreover, from the results (\ref{PsiaaToPsiss}) for $\Psi^{\sigma\sigma'}_{h}(\chi)$ we find  
\begin{align}
\Psi_{h}^{as}(\chi) = 0, \quad  h= 2,4,6,\dots; \quad \Psi_{h}^{sa}(\chi) = 0, \quad h= 1,3,5,\dots\,. \label{PsiDiscr}
\end{align}
Writing explicitly the sums over $h$ defined in (\ref{sumsofh}) and the normalization constants (\ref{NormConst}) in the four-point function (\ref{4ptFinal1}), and using (\ref{KcDiscr}) together with (\ref{PsiDiscr}), we obtain:
\begin{equation}
\begin{aligned}
\frac{\mathcal{F}^{\sigma \sigma'}_{c}(\tau_{1},\tau_{2};\tau_{3},\tau_{4})}{G_{c}(\tau_{12})G_{c}(\tau_{34})}= &\alpha_{0}^{\sigma \sigma'} \int_{1/2+i0}^{1/2+i\infty}\frac{dh}{2\pi i} \frac{(h-1/2)}{2\pi \tan(\pi h)}\left(\frac{K_{c}(h)}{\textbf{1}-K_{c}(h)}\right)^{\sigma \sigma'}\Psi^{\sigma\sigma'}_{h}(\chi) \\
&+ \delta^{\sigma \sigma'}\alpha_{0}^{\sigma \sigma}\sum_{n=1}^{\infty}
\left[\frac{(h-1/2)}{2\pi^2}\frac{k_{\sigma}(h)}{1-k_{\sigma}(h)}\Psi^{\sigma\sigma}_{h}(\chi)\right]_{h=2n+1-\sigma}\,,
\end{aligned}
\end{equation}
where $\sigma,\sigma'=a,s$ (or, when numeric, $=1,2$). 

Using that $\Psi_{1-h}^{\sigma\sigma'}(\chi)=\Psi^{\sigma'\sigma}_{h}(\chi)$
and  (\ref{RelKc1}),  
we obtain a useful relation: 
\begin{equation}
\begin{aligned}
&\alpha_{0}^{\sigma \sigma'} \int_{1/2+i0}^{1/2+i\infty}\frac{dh}{2\pi i} \frac{(h-1/2)}{2\pi \tan(\pi h)}\left(\frac{K_{c}(h)}{\textbf{1}-K_{c}(h)}\right)^{\sigma \sigma'}\Psi^{\sigma\sigma'}_{h}(\chi)  \\
&\qquad\qquad  = \alpha_{0}^{\sigma' \sigma} \int_{1/2-i\infty}^{1/2+i0}\frac{dh}{2\pi i} \frac{(h-1/2)}{2\pi \tan(\pi h)}\left(\frac{K_{c}(h)}{\textbf{1}-K_{c}(h)}\right)^{\sigma' \sigma}\Psi^{\sigma'\sigma}_{h}(\chi)\,, \label{hTo1mheq}
\end{aligned}
\end{equation}
where we also used that $\alpha_{0}^{as}=-(q-1)\alpha_{0}^{sa}$ 
in the domain $0<\chi<1$, since  $\alpha_{0}^{as}/\alpha_{0}^{sa}=-(q-1)\sgn(\chi)\sgn(1-\chi)$. In the domain $1<\chi < 2$, we have $\Psi_{h}^{as}(\chi)=\Psi^{sa}_{h}(\chi)=0$ and thus (\ref{hTo1mheq}) still holds. Using (\ref{hTo1mheq}), we can write the four-point function in the form 
\begin{equation}
\begin{aligned}
\frac{\mathcal{F}_{c}(\tau_{1},\tau_{2};\tau_{3},\tau_{4})}{G_{c}(\tau_{12})G_{c}(\tau_{34})}= &\sum_{\sigma,\sigma'=a,s}\alpha_{0}^{\sigma \sigma'} \int_{1/2-i\infty}^{1/2+i\infty}\frac{dh}{2\pi i} \frac{(h-1/2)}{4\pi \tan(\pi h)}\left(\frac{K_{c}(h)}{\textbf{1}-K_{c}(h)}\right)^{\sigma \sigma'}\Psi^{\sigma\sigma'}_{h}(\chi) \\
&+\sum_{\sigma= a,s}\alpha_{0}^{\sigma \sigma}\sum_{n=1}^{\infty}\left[\frac{(h-1/2)}{2\pi^2}\frac{k_{\sigma}(h)}{1-k_{\sigma}(h)}\Psi^{\sigma\sigma}_{h}(\chi)\right]_{h=2n+1-\sigma}\,. \label{4ptFinal2}
\end{aligned}
\end{equation}
There is a well-known issue with this formula due to divergences associated with the soft modes at $h=1$ and $h=2$, for which $k_{1}(2)=1$ and $k_{2}(1)=1$ for an arbitrary asymmetry parameter $\theta$ (or $\mathscr{E}$) \cite{Maldacena:2016hyu, Klebanov:2016xxf, Gu:2019jub, Tikhanovskaya:2020elb}. To regularize these divergences, one has to compute the $1/(\beta J)$ corrections to the conformal kernel eigenvalues $k_{1}(h)$ and $k_{2}(h)$. This has been done only for $\theta=0$ in \cite{Maldacena:2016hyu, Gu:2019jub}.
Below, we assume that we work with the four-point function in which the $h=1$ and $h=2$ modes are excluded.

Now we focus on the domain $0<\chi < 1$. Using the formulas for $\Psi^{\sigma \sigma'}_{h}(\chi)$ in (\ref{PsiaaToPsiss}) for this domain and making transformation $h\to 1-h$ in different terms of the four-point function, we can bring it to the form 
\begin{equation}
\begin{aligned}
&\frac{\mathcal{F}_{c}(\tau_{1},\tau_{2};\tau_{3},\tau_{4})}{G_{c}(\tau_{12})G_{c}(\tau_{34})}= \sum_{\sigma=a,s}\bigg[\alpha_{0}^{\sigma a} \int_{\mathcal{C}^{a}}\frac{dh}{2\pi i} \frac{(h-1/2)}{2\pi \tan(\pi h /2)}\left(\frac{K_{c}(h)}{\textbf{1}-K_{c}(h)}\right)^{\sigma a} \\
&\qquad +\alpha_{0}^{\sigma s} \int_{\mathcal{C}^{s}}\frac{dh}{2\pi i} \frac{(h-1/2)}{2\pi \cot(\pi h/2)}\left(\frac{K_{c}(h)}{\textbf{1}-K_{c}(h)}\right)^{\sigma s}\bigg]\frac{\Gamma(h)^{2}}{\Gamma(2h)} \chi^h\,_2F_1(h,h;2h;\chi)\,,
\label{4ptFinal3}
\end{aligned}
\end{equation}
where we introduced two contours: $\mathcal{C}^{a}$ and $\mathcal{C}^{s}$ in the complex $h$ plane as 
\begin{equation}
\begin{aligned}
\int_{\mathcal{C}^{a}}\frac{dh}{2\pi i} = \int_{1/2-i\infty}^{1/2+i\infty}\frac{dh}{2\pi i} + \sum_{n=2}^{\infty} \operatorname*{Res}_{h=2n}, \quad 
\int_{\mathcal{C}^{s}}\frac{dh}{2\pi i} = \int_{1/2-i\infty}^{1/2+i\infty}\frac{dh}{2\pi i} + \sum_{n=1}^{\infty} \operatorname*{Res}_{h=2n+1} \,,
\end{aligned}
\end{equation}
and now the sums over integer values of $h$ are written as sums over residues of the poles of $1/\tan(\pi h/2)$ and $1/\cot( \pi h/2)$.  In deriving (\ref{4ptFinal3}), we again used relations (\ref{RelKc1}) and (\ref{KcDiscr}), together with $\alpha_{0}^{as}=-(q-1)\alpha_{0}^{sa}$.

\begin{figure}[h!]
    \centering
    \includegraphics[width=7cm]{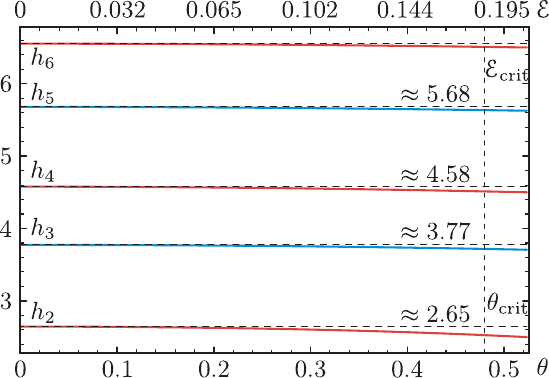}
    \caption{Lowest scaling dimensions of the $q=4$ cSYK operators as functions of $\theta$ and $\mathscr{E}$. The horizontal dashed lines indicate their values at $\theta=0$ ($\mathscr{E}=0$) and the vertical dashed line marks the critical value $\theta_{\textrm{crit}}\approx 0.48$ ($\mathscr{E}_{\textrm{crit}}\approx 0.18$), at which the cSYK model undergoes the first-order phase transition.}
    \label{fig::anomalous_dim}
\end{figure}

The matrix elements 
\begin{equation}
\left(\frac{K_{c}(h)}{\textbf{1}-K_{c}(h)}\right)^{\sigma \sigma'} = \frac{k_{1}(h)}{1-k_{1}(h)}v_{1}^{\sigma}w_{1}^{\sigma'} +
\frac{k_{2}(h)}{1-k_{2}(h)}v_{2}^{\sigma}w_{2}^{\sigma'} \label{Kc1Kc}
\end{equation}
have poles at $h= h_{m}$, $m=0,1,2,\dots$, where $h_{m}$ are the scaling dimensions of the bilinear operators $\mathcal{O}_{m}(\tau) = \bar{\psi}_{i}\partial_{\tau}^{m}\psi^{i}$ and they satisfy\footnote{We note that for $q=4$ these formulas are valid only for $|\theta|\leq \pi/6$ \cite{Gu:2019jub}, but this is already beyond the critical value $|\theta_{\textrm{crit}}|\approx 0.153\pi$, above which the conformal phase is unstable and the cSYK model is in the “low-entropy” phase \cite{Azeyanagi_2018, Ferrari:2019ogc, Tikhanovskaya:2020elb}. } 
\begin{equation}
k_{1}(h_{2m+1}) = 1, \quad k_{2}(h_{2m})=1, \quad m= 0,1,2\dots\,. \label{AnDims}
\end{equation}
At $\theta = 0$ ($\mathscr{E}=0$) the first scaling dimensions are $h_{0}=1$, $h_{1}=2$, $h_{2}\approx 2.65$, $h_{3}\approx 3.77$, $h_{4}\approx 4.58$ and $h_{5}\approx 5.68$ and $\mathcal{O}_{0}$ and $\mathcal{O}_{1}$, as we mentioned in the introduction, correspond to the conserved charge density and the Hamiltonian of the cSYK model and are responsible for the breaking of the conformal symmetry. The scaling dimensions of the other operators depend on the asymmetry parameter and their dependence on $\theta$ (and $\mathscr{E}$)  for $q=4$ is shown in Figure~\ref{fig::anomalous_dim}. 
We note that although the scaling dimensions depend on the asymmetry parameter, their deviation from the $\theta=0$ values is small. This is usually not the case for other SYK type models \cite{Tikhanovskaya:2020elb,Tikhanovskaya:2020zcw}.

Now,  by deforming the contours $\mathcal{C}^{a}$ and $\mathcal{C}^{s}$ in (\ref{4ptFinal3}) to pick up the poles of the matrix elements in (\ref{Kc1Kc}), we find
\begin{equation}
\begin{aligned}
\frac{\mathcal{F}_{c}(\tau_{1},\tau_{2};\tau_{3},\tau_{4})}{G_{c}(\tau_{12})G_{c}(\tau_{34})}=& \sum_{\sigma=a,s} \sum_{h=h_{m}} \bigg[ \alpha_{0}^{\sigma a} \frac{(h-1/2)}{2\pi \tan(\pi h/2)} \Big(\frac{1}{K_{c}'(h)}\Big)^{\sigma a}  \\
& \qquad +\alpha_{0}^{\sigma s} \frac{(h-1/2)}{2\pi \cot(\pi h/2)} \Big(\frac{1}{K_{c}'(h)}\Big)^{\sigma s}\bigg]\frac{\Gamma(h)^2}{\Gamma(2h)}\chi^h \,_2F_1(h,h;2h;\chi)\,, \label{4ptFinal4}
\end{aligned}
\end{equation}
where we have excluded $h_{0}=1$ and $h_{1}=2$ from this sum, and have also defined
\begin{equation}
\left(\frac{1}{K_{c}'(h)}\right)^{\sigma \sigma'} \equiv 
\begin{cases}
\frac{v_{1}^{\sigma}w_{1}^{\sigma'}}{k'_{1}(h)},\quad  h = h_{2m+1} \\
\frac{v_{2}^{\sigma}w_{2}^{\sigma'}}{k'_{2}(h)}, \quad h = h_{2m} 
\end{cases}\,, \quad m =1,2,3,\dots\,.
\end{equation}
As we show in the next section, this expression is expected from the CFT analysis of the cSYK model, where $\chi^{h_{m}} \,_2F_1(h_{m},h_{m};2h_{m};\chi)$ are the conformal blocks and the prefactors correspond to the structure constants.

\section{CFT approach to the four-point function and OPE data}
\label{4ptCFTandStr}
In this section we compute the four-point function, assuming that in the large $N$ limit the complex SYK model is described by NCFT$_{1}$ in the strong coupling regime $\beta J \gg 1$. The conformal symmetry fixes the form of the two and three-point functions of the operators, up to structure constants. Using this, we deduce the operator product expansion (OPE) and apply it to compute the four-point function. 
For simplicity, we work at zero temperature. 

The three-point function between two fermionic fields $\psi^i$ and $\psibar_i$ and a scalar operator $\mathcal{O}_{h}$ in CFT$_{1}$ reads \cite{Nobili:1973yu, Iliesiu:2015akf, Tikhanovskaya:2020elb}
\begin{align}
    \frac{1}{N}\langle \psi^{i}(\tau_{1})\psibar_{i}(\tau_{2})\mathcal{O}_{h}(\tau_0)\rangle=  G_{c}(\tau_{12}) \frac{c_{h}^{a}+c_{h}^{s} \sgn(\tau_{12}) \sgn(\tau_{10})\sgn(\tau_{20})}{|\tau_{12}|^{-h}|\tau_{10}|^{h}|\tau_{20}|^{h}}\,,
    \label{eq_3pt_function}
\end{align}
where the sum over $i$ is assumed and we introduced the structure constants $c^{a}_{h}$ and $c^{s}_{h}$. 

We can write the four-point function by means of the operator product expansion (OPE) for $\psi^i$ and $\psibar_i$, inferred from the three-point function (\ref{eq_3pt_function}):
\begin{equation}
\begin{aligned}
    \frac{1}{N}\psi^{i}(\tau_1) &\psibar_{i}(\tau_2)= G_{c}(\tau_{12}) +  \\
    & + \frac{1}{N}G_{c}(\tau_{12})\sum_{h=h_{m}} \Big( \frac{c_{h}^{a}}{|\tau_{12}|^{-h}}C^{a}_{h}(\tau_{12},\partial_2) +  \frac{c_{h}^{s}\sgn(\tau_{12})}{|\tau_{12}|^{-h}}C^{s}_{h}(\tau_{12},\partial_2) \Big)\mathcal{O}_{h}(\tau_{2})\,,
    \label{ccOPE}
\end{aligned}
\end{equation}
where $C^{a}_{h}$ and $C^{s}_{h}$ are the generators of the descendant operators of the theory. Applying this OPE, and using $\langle \mathcal{O}_h(\tau) \mathcal{O}_{h'}(0)\rangle = N\delta_{hh'}/|\tau|^{2h}$ for the two-point function\footnote{Note that the $\mathcal{O}_h$s are conventionally \cite[(3.12)]{Tikhanovskaya:2020elb} taken to have this propagator, rather than a unit-normalized one: this avoids factors of $\sqrt{N}$ in the equations.}, we obtain 
\begin{equation}
\begin{aligned}
    &\frac{1}{N^2}\langle \psi^i(\tau_1)  \psibar_i(\tau_2)  \psi^j(\tau_3)\psibar_j(\tau_4)\rangle = G_{c}(\tau_{12})G_{c}(\tau_{34}) + \frac{G_{c}(\tau_{12})G_{c}(\tau_{34})}{N} \sum_{h=h_{m}} |\tau_{12}|^{h}|\tau_{34}|^{h} \\
    & \quad \Big((c_{h}^{a})^2C_{h}^{a}(\tau_{12},\partial_2) C_{h}^{a}(\tau_{34},\partial_4) + c_{h}^{a}c_{h}^{s} \, \sgn(\tau_{34}) C_{h}^{a}(\tau_{12},\partial_2) C_{h}^{s}(\tau_{34},\partial_4) +  \\
    &  c_{h}^{s}c_{h}^{a} \,\sgn(\tau_{12}) C_{h}^{s}(\tau_{12},\partial_2) C_{h}^{a}(\tau_{34},\partial_4) +  (c_{h}^{s})^2 \sgn(\tau_{12}\tau_{34}) C_{h}^{s}(\tau_{12},\partial_2) C_{h}^{s}(\tau_{34},\partial_4)\Big) \frac{1}{|\tau_{24}|^{2h}},
\end{aligned}
\end{equation}
where the sum over indices $i$ and $j$ is assumed. 
By performing computations with 
$C_{h}^{a}(\tau_{12},\partial_2)$ and $C_{h}^{s}(\tau_{12},\partial_2)$ discussed in Appendix~\ref{AppCompOPE}, we find for the connected part of the four-point function 
\begin{equation}
\begin{aligned}
\frac{\mathcal{F}_{c}(\tau_{1},\tau_{2};\tau_{3},\tau_{4})}{G_{c}(\tau_{12})G_{c}(\tau_{34})} =& \sum_{h=h_{m}} \Big((c_{h}^{a})^2    
     + c_{h}^{a}c_{h}^{s} \,\sgn(\tau_{34}\tau_{13}\tau_{14}) \\
&+ c_{h}^{s}c_{h}^{a} \,\sgn(\tau_{12}\tau_{13}\tau_{23})  + (c_{h}^{s})^2 \sgn(\chi) \Big) |\chi|^{h}\,_2F_1(h,h;2h;\chi)\,.
    \label{conformal_4ptMain}
\end{aligned}
\end{equation}
We also provide an alternate derivation of this equation by means of the shadow formalism in Appendix~\ref{app:shadowformalism}.

Comparing this result with \eqref{4ptFinal4}
we read off products of the structure constants $(c^a_{h})^2,\ c^a_{h}c^s_{h},\ c^s_{h}c^a_{h},$ and $(c^s_{h})^2$, for the  scaling dimensions $h=h_{m}$, $m=2,3,4,\dots$:
\begin{equation}
\begin{aligned}
(c^{a}_{h})^2 & = \frac{1}{b(q-1)} \frac{(h-1/2)}{2 \pi \tan(\pi h /2)}  \frac{\Gamma (h)^2}{\Gamma (2h)}\left(\frac{1}{K'_{c}(h)}\right)^{aa} \,, \\
c^{s}_{h}c^{a}_{h} & = \frac{1}{b(q-1)} \frac{(h-1/2)}{2 \pi \tan(\pi h /2)}   \frac{\Gamma (h)^2}{\Gamma (2h)} \left(\frac{1}{K'_{c}(h)}\right)^{sa}\,, \\
c^{a}_{h}c^{s}_{h} & = -\frac{1}{b} \frac{(h-1/2)}{2 \pi \cot(\pi h /2)} \frac{\Gamma (h)^2}{\Gamma (2h)} \left(\frac{1}{K'_{c}(h)}\right)^{as} \,, \\
(c^{s}_{h})^2 & = -\frac{1}{b} \frac{(h-1/2)}{2 \pi \cot(\pi h /2)}  \frac{\Gamma (h)^2}{\Gamma (2h)} \left(\frac{1}{K'_{c}(h)}\right)^{ss}\,. \label{strcFinal}
\end{aligned}
\end{equation}
To check the consistency of this result, we note that the components of the  left and right eigenfunctions of $K_{c}(h)$ in (\ref{v1v2right}) and (\ref{w1w2left}) obey, for any $h$, $\theta$, and $\Delta$:
\begin{align}
\frac{1}{q-1}\frac{v_{i}^{s}w_{i}^{a}}{\tan(\pi h /2)} = -\frac{v_{i}^{a}w_{i}^{s}}{\cot(\pi h /2)}, \quad i = 1,2\,,
\end{align}
and therefore, the expressions in (\ref{strcFinal}) indeed satisfy
\begin{align}
c_{h}^{s}c_{h}^{a} = c_{h}^{a}c_{h}^{s}\,.
\end{align}
Similarly, one can easily verify that  $(c^{a}_{h})^2 (c^{s}_{h})^2 = (c^{a}_{h}c^{s}_{h}) (c^{s}_{h}c^{a}_{h})$, as expected. We also note that $(c_{h}^{a})^2,\, (c_{h}^{s})^2>0$ and $c_{h}^{a}c_{h}^{s}<0$, for all $h=h_{m}$. Thus, without loss of generality, we can choose $c_{h}^{a}>0$ and $c_{h}^{s}<0$.

\begin{figure}[h!]
    \centering
    \includegraphics[width=7cm]{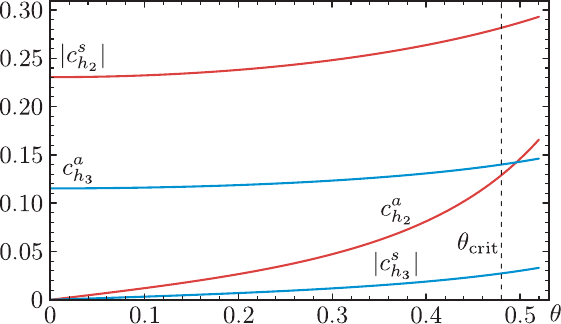}
    \caption{The structure constants $c_{h}^{a}$ and $c_{h}^{s}$ for the first two scaling dimensions $h_{2}$ and $h_{3}$ as functions of $\theta$ for $q=4$ cSYK model. The vertical dashed line marks the critical value $\theta_{\textrm{crit}}\approx 0.48$, at which the cSYK model undergoes the first-order phase transition.}
    \label{fig::structures}
\end{figure}

At $\mathscr{E}=0$ ($\theta = 0$), the only non-zero structure constants are
\begin{equation}
\begin{aligned}
    (c^{a}_{h})^2 & = \frac{1}{b(q-1)} \frac{(h-1/2)}{2 \pi \tan(\pi h /2)} \frac{1}{(k^{a}_{c}(h))'}\frac{\Gamma(h)^2}{\Gamma (2h)}\,, \quad h= h_{2m+1}\,,\\
    (c^{s}_{h})^2 & = -\frac{1}{b} \frac{(h-1/2)}{2 \pi \cot(\pi h /2)} \frac{1}{(k^{s}_{c}(h))'}\frac{\Gamma (h)^2}{\Gamma (2h)}\,, \quad h =h_{2m}\,,
\end{aligned}
\end{equation}
where $m=1,2,3,\dots$, $k_{c}^{a}(h)$ and $k_{c}^{s}(h)$ are given in (\ref{kacksc}), and the result for $(c^{a}_{h})^{2}$ coincides with \cite{Maldacena:2016hyu} (eq. (3.92))\footnote{We stress that our result differs by an extra factor of $1/2$ because in our case $N$ denotes the number of complex fermions, whereas in \cite{Maldacena:2016hyu} $N$ denotes the number of Majorana fermions.}. 
Finally, as an example, in Figure~\ref{fig::structures} we plot the dependence of the structure constants $c_{h}^{a}$ and $c_{h}^{s}$ on the asymmetry parameter $\theta$ for the first two scaling dimensions, $h_{2}$ and $h_{3}$, in the $q=4$ case.

\section{Concluding Remarks}

In this paper, we computed the complex SYK model four-point function in the conformal limit for an arbitrary value of the asymmetry parameter $\mathscr{E}$ (or $\theta$) at leading order in $1/N$ using two approaches. The first one uses a direct resummation of the ladder diagrams in the large $N$ limit, and subsequent use of the conformal limit. This result for the four-point function is presented in (\ref{4ptFinal2}). The second approach utilizes conformal symmetry of the cSYK model in the same limits and provides the result for the four-point function in terms of the sums over conformal blocks with unknown structure constants (\ref{conformal_4ptMain}). We found full agreement between the two computations, which allowed us to obtain explicit expressions for the structure constants, listed in (\ref{strcFinal}). For a non-zero asymmetry parameter, all of these structure constants become non-zero and contribute to various observables of the cSYK model.

It is well known that SYK models contain soft modes \cite{Maldacena:2016hyu, Jevicki:2016bwu, Kitaev:2017awl, Gu:2019jub}, which break conformal symmetry and lead to divergences in the conformal four-point function, as we mentioned below (\ref{4ptFinal2}). In the conformal limit, these modes must be treated separately.  The large $q$ limit, on the other hand, allows one to solve the large $N$ SYK model at all scales, free of the soft modes divergences, and provides a direct interpolation between UV and IR regimes \cite{Maldacena:2016hyu, Tarnopolsky:2018env, Tikhanovskaya:2020elb}. It is also possible to solve the SYK model in the double scaling limit, in which both $q$ and $N$ are taken to infinity while $\lambda \equiv q^{2}/N$ is kept fixed, realizing the Double-Scaled SYK (DSSYK) model \cite{Berkooz:2018qkz, Berkooz:2018jqr, Berkooz:2024lgq}. Recently, a large $q$ approach to the complex SYK model with a non-zero asymmetry parameter  was developed \cite{Gubankova:2025gbx}, and agreement with the double-scaled complex SYK model \cite{Berkooz:2020uly} was established for the grand potential and the two-point function in the $\lambda \to 0$ limit. It would be interesting to derive the large $q$ four-point function of the cSYK model, in analogy with the corresponding results for the Majorana SYK model \cite{Streicher:2019wek, Choi:2019bmd}.

We conclude by noting that the complex SYK model in the presence of a chemical potential acquires new and interesting properties that are not present in the Majorana SYK case. The connection between the Majorana SYK model and models of quantum gravity, black hole physics, and chaos \cite{Maldacena:2016upp, Cotler:2016fpe, Engelsoy:2016xyb, Roberts:2018mnp, Qi:2018bje, Saad:2019lba, Orman:2024mpw, Dodelson:2024atp, Dodelson:2025rng} makes it appealing to explore these relations further, now including the additional features that arise in the cSYK model at non-zero chemical potential.

\subsection*{Acknowledgements}
We thank Igor R. Klebanov and Subir Sachdev for stimulating discussions. We also thank  Igor R. Klebanov for comments on the draft. E.A.C, L. F-T and G. T. were supported  by the U.S. Department of Energy Grant
No. DE-SC0010118. 

\appendix

\section{\texorpdfstring{Action of the kernel $K_{c}$ and $\mathcal{F}_{0}$ on the basis functions $W_{h}^{\sigma}$ and $\tilde{W}_{h}^{\sigma}$}{Action of the kernel K and F on the basis functions W}}
\label{AppDiagcSYK}
\begin{figure}[h!]
\centering
\includegraphics[width=11.5cm]{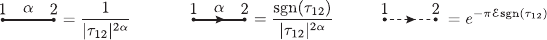}
\caption{Diagrammatic representation of the basic propagators. We introduced a separate dashed line, which represents the asymmetry factor, and it is assumed that it is always $\mathscr{E}$. For $\mathscr{E}=0$ the asymmetry propagator disappears.}
\label{props}
\end{figure}

\begin{figure}[h!]
\centering
\includegraphics[width=10cm]{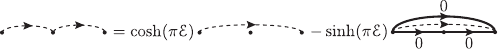}
\caption{Diagrammatic representation of the identity (\ref{AssymPropRel}).}
\label{props_ids2}
\end{figure}

In this section, we work at zero-temperature $\beta = \infty$. The results obtained in this regime coincide with those at finite temperature due to the reparameterization invariance of the cSYK model in the conformal limit.
We start by introducing  in Figure \ref{props} diagrammatic representations for basic propagators in the complex SYK model with non-zero asymmetry parameter $\mathscr{E}$ (non-zero chemical potential $\mu$). 
We can derive an important identity for the asymmetry propagator, which we will use frequently below. Namely, we have 
\begin{equation}
\begin{aligned}
e^{-\pi \mathscr{E} \sgn(\tau_{12})} e^{-\pi \mathscr{E} \sgn(\tau_{23})} &= e^{-\pi \mathscr{E} \sgn(\tau_{13})} e^{-\pi \mathscr{E} \sgn(\tau_{13}\tau_{12}\tau_{23}) } \\
&= 
 e^{-\pi \mathscr{E} \sgn(\tau_{13})} (\cosh(\pi \mathscr{E}) -\sinh( \pi \mathscr{E})\sgn(\tau_{13}\tau_{12}\tau_{23}))\,, \label{AssymPropRel}
\end{aligned}
\end{equation}
where we used the identity $\sgn(a)+\sgn(b)= \sgn(a+b)(1+\sgn(ab))$. The identity (\ref{AssymPropRel}) is represented diagrammatically in Figure \ref{props_ids2}.

\begin{figure}[h!]
\centering
\includegraphics[width=11cm]{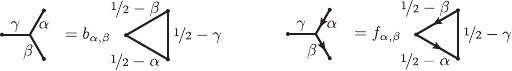}
\caption{Diagrammatic representation of the star-triangle identities, where $\alpha+\beta+\gamma=1$ and the coefficients $b_{\alpha,\beta}$ and $f_{\alpha,\beta}$ are given in  (\ref{cofsfb}). }
\label{star_triag_ids}
\end{figure}

The star-triangle identities read 
\begin{equation}
\begin{aligned}
\int_{-\infty}^{+\infty}\frac{d\tau_{0}}{|\tau_{10}|^{2\alpha}|\tau_{20}|^{2\beta}|\tau_{30}|^{2\gamma}}= \frac{b_{\alpha,\beta}}{|\tau^{2}_{12}|^{1/2-\gamma}|\tau^{2}_{23}|^{1/2-\alpha}|\tau^{2}_{13}|^{1/2-\beta}}\,, \\
\int_{-\infty}^{+\infty}\frac{d\tau_{0} \sgn(\tau_{10})\sgn(\tau_{02})}{|\tau_{10}|^{2\alpha}|\tau_{20}|^{2\beta}|\tau_{30}|^{2\gamma}}= \frac{f_{\alpha,\beta} \sgn(\tau_{13})\sgn(\tau_{32})}{|\tau^{2}_{12}|^{1/2-\gamma}|\tau^{2}_{23}|^{1/2-\alpha}|\tau^{2}_{13}|^{1/2-\beta}}\,, \label{st_tr_ids}
\end{aligned}
\end{equation}
where $\alpha+\beta+\gamma=1$ and the coefficients $f_{\alpha,\beta}$
and $b_{\alpha,\beta}$ are 
\begin{align}
b_{\alpha,\beta}=\sqrt{\pi}\frac{\Gamma(\frac{1}{2}-\alpha)
\Gamma(\frac{1}{2}-\beta)\Gamma(\frac{1}{2}-\gamma)}{\Gamma(\alpha)\Gamma(\beta)\Gamma(\gamma)}\,, \quad f_{\alpha,\beta}=\sqrt{\pi}\frac{\Gamma(1-\alpha)
\Gamma(1-\beta)\Gamma(\frac{1}{2}-\gamma)}{\Gamma(\frac{1}{2}+\alpha)\Gamma(\frac{1}{2}+\beta)\Gamma(\gamma)}\,. \label{cofsfb}
\end{align}
The star-triangle identities are represented diagrammatically in Figure \ref{star_triag_ids}.

\begin{figure}[H]
\centering
\includegraphics[width=12cm]{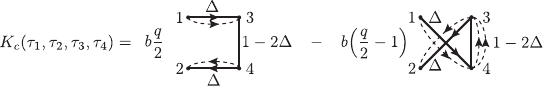}
\caption{Diagrammatic representation of the conformal kernel $K_{c}$ in (\ref{KcT0}).}
\label{Kc_diag}
\end{figure}

\begin{figure}[h!]
\centering
\includegraphics[width=12cm]{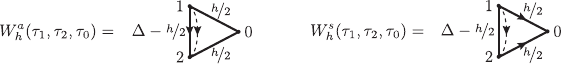}
\caption{Diagrammatic representation of the right basis functions $W_{h}^{a}$ and $W_{h}^{s}$.}
\label{WaWs_diags}
\end{figure}

\begin{figure}[h!]
\centering
\includegraphics[width=10cm]{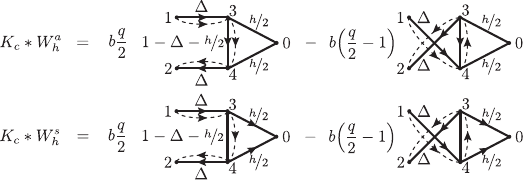}
\caption{Diagrammatic representation of the action of the conformal kernel $K_{c}$ on the right basis functions $W_{h}^{a}$ and $W_{h}^{s}$.}
\label{Kc_WaWs_diags}
\end{figure}

\begin{figure}[h!]
\centering
\includegraphics[width=12cm]{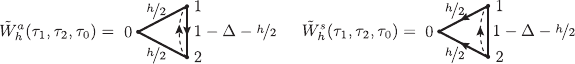}
\caption{Diagrammatic representation of the left basis functions $\tilde{W}_{h}^{a}$ and $\tilde{W}_{h}^{s}$.}
\label{WtaWts_diags}
\end{figure}

\begin{figure}[h!]
\centering
\includegraphics[width=11cm]{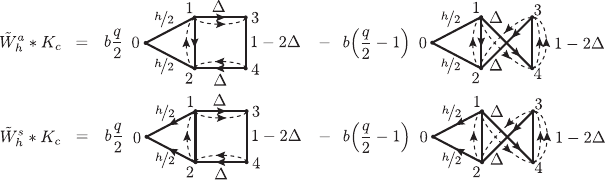}
\caption{Diagrammatic representation of the action of the conformal kernel $K_{c}$ on the left basis functions $\tilde{W}_{h}^{a}$ and $\tilde{W}_{h}^{s}$. The dimension of the middle vertical line in all diagrams is  $1-\Delta-h/2$.}
\label{Kc_WtaWts_diags}
\end{figure}

Now we are ready to apply this diagrammatic technique to the computation of the action of the conformal kernel $K_{c}$ and the function $\mathcal{F}_{0}$ to the left and right basis functions $\tilde{W}_{h}^{\sigma}$ and $W_{h}^{\sigma}$ with $\sigma=a,s$. 
The kernel (\ref{kernel}) in the conformal limit and at zero temperature can be written in the form 
\begin{equation}
\begin{aligned}
K_{c}(\tau_{1},\tau_{2};\tau_{3},\tau_{4})  = & \; b\ \frac{q}{2}\frac{\sgn(\tau_{13})\sgn(\tau_{42}) }{|\tau_{13}|^{2\Delta}|\tau_{24}|^{2\Delta}|\tau_{34}|^{2-4\Delta}}e^{-\pi \mathscr{E}(\sgn(\tau_{13})+\sgn(\tau_{42}))} \\
&- b\left(\frac{q}{2} -1 \right) \frac{\sgn(\tau_{14})\sgn(\tau_{32}) e^{-\pi \mathscr{E}(\sgn(\tau_{14})+\sgn(\tau_{32})+2\sgn(\tau_{43}))}}{|\tau_{14}|^{2\Delta}|\tau_{32}|^{2\Delta}|\tau_{34}|^{2-4\Delta}}\,,
\label{KcT0}
\end{aligned}
\end{equation}
where we used that the conformal Green's function (\ref{GcSol}) at zero temperature is 
\begin{align}
G_{c}(\tau) = b^{\Delta}\frac{\sgn(\tau)}{|J\tau|^{2\Delta}}e^{-\pi \mathscr{E}\sgn(\tau)}\,, \label{GcAtT0}
\end{align}
and the constant $b$ is given in (\ref{Constb}).
We represent the kernel diagrammatically in Figure \ref{Kc_diag}.
The right basis functions $W_{h}^{a}$ and $W_{h}^{s}$  defined in (\ref{WhaWhs}) read in the zero-temperature limit:
\begin{align}
W^{a}_{h}(\tau_{1},\tau_{2};\tau_{0}) = \frac{\sgn(\tau_{12}) e^{-\pi \mathscr{E} \sgn(\tau_{12})}}{|\tau_{12}|^{2\Delta-h}|\tau_{10}|^{h}|\tau_{20}|^{h}}, \quad 
W^{s}_{h}(\tau_{1},\tau_{2};\tau_{0}) = \frac{\sgn(\tau_{10})\sgn(\tau_{20})e^{-\pi \mathscr{E} \sgn(\tau_{12})}}{|\tau_{12}|^{2\Delta-h}|\tau_{10}|^{h}|\tau_{20}|^{h}}\,. \label{WaWsT0}
\end{align}
and they are represented diagrammatically in Figure  \ref{WaWs_diags}.
The action of the kernel on these basis functions is represented in Figure \ref{Kc_WaWs_diags}, where we defined $K_{c}*W_{h}^{\sigma} \equiv \int d\tau_{3}d\tau_{4}K_{c}(\tau_{1},\tau_{2};\tau_{3},\tau_{4})W_{h}^{\sigma}(\tau_{3},\tau_{4};\tau_{0})$ for $\sigma=a,s$.
Using the star-triangle identities (\ref{st_tr_ids}) and the identity (\ref{AssymPropRel}) we find 
\begin{align}
\begin{cases}
K_{c} * W^{a}_{h} = W_{h}^{a}(\tau_{1},\tau_{2};\tau_{0}) K_{h}^{aa} + W_{h}^{s}(\tau_{1},\tau_{2};\tau_{0})K^{sa}_{h} \\
K_{c} * W^{s}_{h} = W_{h}^{a}(\tau_{1},\tau_{2};\tau_{0})K_{h}^{as} + W_{h}^{s}(\tau_{1},\tau_{2};\tau_{0})K^{ss}_{h}
\end{cases}\,,
\end{align}
where the coefficients $K_{h}^{\sigma\sigma'}$ with $\sigma, \sigma'=a,s$ read
\begin{equation}
\begin{aligned}
&K_{h}^{aa}= -b(q-1)\big(\cosh^{2}(\pi \mathscr{E}) f_{\Delta,1-\Delta-\frac{h}{2}}f_{\Delta,\frac{1+h}{2}-\Delta} - \sinh^{2}(\pi \mathscr{E})b_{\Delta,1-\Delta-\frac{h}{2}}b_{\Delta,\frac{1-h}{2}}\big)\,, \\
&K_{h}^{sa}= b(q-1)\cosh(\pi \mathscr{E})\sinh(\pi \mathscr{E}) \big(f_{\Delta, 1-\Delta-\frac{h}{2}}f_{\frac{1-h}{2},\frac{1+h}{2}-\Delta} +b_{\Delta,1-\Delta-\frac{h}{2}}f_{\frac{1-h}{2},\Delta} \big)\,, \\
&K_{h}^{as}= -b\cosh(\pi \mathscr{E})\sinh(\pi \mathscr{E}) \big(f_{\Delta,\frac{h}{2}}b_{\Delta,\frac{1-h}{2}} +f_{\frac{h}{2},1-\Delta-\frac{h}{2}}f_{\Delta,\frac{1+h}{2}-\Delta} \big)\,, \\
&K_{h}^{ss}=-b(\cosh^{2}(\pi \mathscr{E}) f_{\Delta,\frac{h}{2}} f_{\Delta, \frac{1-h}{2}}-\sinh^{2}(\pi \mathscr{E}) f_{\frac{h}{2},1-\Delta-\frac{h}{2}}f_{\frac{1-h}{2},\frac{1+h}{2}-\Delta} \big)\,,\label{KssprRes}
\end{aligned}
\end{equation}
with $b_{\alpha,\beta}$ and $f_{\alpha,\beta}$  given in (\ref{cofsfb}).
The left basis functions  $\tilde{W}_{h}^{a}$ and $\tilde{W}_{h}^{s}$  defined in (\ref{WthaWths}) read in the zero-temperature limit:
\begin{align}
\tilde{W}^{a}_{h}(\tau_{1},\tau_{2};\tau_{0}) = \frac{\sgn(\tau_{12}) e^{-\pi \mathscr{E} \sgn(\tau_{21})}}{|\tau_{12}|^{2-2\Delta-h}|\tau_{10}|^{h}|\tau_{20}|^{h}}, \quad 
\tilde{W}^{s}_{h}(\tau_{1},\tau_{2};\tau_{0}) = \frac{\sgn(\tau_{10})\sgn(\tau_{20})e^{-\pi \mathscr{E} \sgn(\tau_{21})}}{|\tau_{12}|^{2-2\Delta-h}|\tau_{10}|^{h}|\tau_{20}|^{h}}\,. \label{WtaWtsT0}
\end{align}
The diagrammatic representation for the left basis functions $\tilde{W}_{h}^{a}$ and $\tilde{W}_{h}^{s}$ is shown in Figure \ref{WtaWts_diags}.
The action of the kernel on these basis functions is represented in Figure \ref{Kc_WtaWts_diags}, where we defined $\tilde{W}_{h}^{\sigma}*K_{c} \equiv \int d\tau_{1}d\tau_{2} \tilde{W}_{h}^{\sigma}(\tau_{1},\tau_{2};\tau_{0})K_{c}(\tau_{1},\tau_{2};\tau_{3},\tau_{4})$ for $\sigma=a,s$.

\noindent Using the star-triangle identities (\ref{st_tr_ids}) and the identity (\ref{AssymPropRel}) we find 
\begin{align}
\begin{cases}
\tilde{W}^{a}_{h}*K_{c}  = \tilde{K}_{h}^{aa}\tilde{W}_{h}^{a}(\tau_{3},\tau_{4};\tau_{0}) + \tilde{K}^{as}_{h}\tilde{W}_{h}^{s}(\tau_{3},\tau_{4};\tau_{0}) \\
\tilde{W}^{s}_{h}*K_{c}  = \tilde{K}_{h}^{sa}\tilde{W}_{h}^{a}(\tau_{3},\tau_{4};\tau_{0}) + \tilde{K}^{ss}_{h}\tilde{W}_{h}^{s}(\tau_{3},\tau_{4};\tau_{0})
\end{cases}\,,
\end{align}
where the coefficients $\tilde{K}_{h}^{\sigma\sigma'}$ with $\sigma, \sigma'=a,s$ are simply related to   $K_{h}^{\sigma\sigma'}$ in (\ref{KssprRes}) as
\begin{align}
\tilde{K}^{\sigma\sigma'}_{h} = K^{\sigma \sigma'}_{1-h} \,.
\end{align}

\begin{figure}[h!]
\centering
\includegraphics[width=12cm]{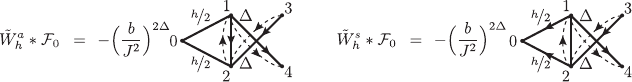}
\caption{Diagrammatic representation of the action of the  $\mathcal{F}_{0}$ on the left basis functions $\tilde{W}_{h}^{a}$ and $\tilde{W}_{h}^{s}$. The dimension of the middle vertical line in both diagrams is  $1-\Delta-h/2$.}
\label{WtF0_diags}
\end{figure}

For the function $\mathcal{F}_{0}=-G(\tau_{14})G(\tau_{32})$ at zero temperature, we find  
\begin{align}
\mathcal{F}_{0}(\tau_{1},\tau_{2};\tau_{3},\tau_{4}) =-\Big(\frac{b}{J^{2}}\Big)^{2\Delta} \frac{\sgn(\tau_{14})\sgn(\tau_{32})}{|\tau_{14}|^{2\Delta}|\tau_{32}|^{2\Delta}}e^{-\pi \mathscr{E}(\sgn(\tau_{14})+\sgn(\tau_{32}))}\,.
\end{align}
We present diagrams for $\tilde{W}^{a}_{h}*\mathcal{F}_{0}$ and $\tilde{W}^{s}_{h}*\mathcal{F}_{0}$ in Figure \ref{WtF0_diags}.
These diagrams are similar to the ones we had for $\tilde{W}^{\sigma}_{h}*K_{c}$ with $\sigma=a,s$ and after similar computations we find  
\begin{align}
\begin{cases}
\tilde{W}_{h}^{a}*\mathcal{F}_{0} = \frac{1}{b}\Big(\frac{b}{J^{2}}\Big)^{2\Delta}\Big(\frac{1}{q-1}\tilde{K}_{h}^{aa}W^{a}_{h}(\tau_{3},\tau_{4};\tau_{0}) -\tilde{K}^{as}_{h}W^{s}_{h}(\tau_{3},\tau_{4};\tau_{0})\Big) \\
\tilde{W}_{h}^{s}*\mathcal{F}_{0}  = \frac{1}{b}\Big(\frac{b}{J^{2}}\Big)^{2\Delta}\Big(\frac{1}{q-1}\tilde{K}_{h}^{sa}W^{a}_{h}(\tau_{3},\tau_{4};\tau_{0}) -
\tilde{K}^{ss}_{h}W^{s}_{h}(\tau_{3},\tau_{4};\tau_{0}) \Big)
\end{cases}\,.
\end{align}
Using that $\tilde{K}^{\sigma\sigma'}_{h}=K^{\sigma\sigma'}_{1-h}$ we finally obtain 
\begin{align}
\begin{cases}
\tilde{W}_{1-h}^{a}*\mathcal{F}_{0}  = \frac{1}{b}\Big(\frac{b}{J^{2}}\Big)^{2\Delta}\Big(\frac{1}{q-1}K_{h}^{aa}W^{a}_{1-h}(\tau_{3},\tau_{4};\tau_{0}) -K^{as}_{h}W^{s}_{1-h}(\tau_{3},\tau_{4};\tau_{0})\Big) \\
\tilde{W}_{1-h}^{s}*\mathcal{F}_{0}  = \frac{1}{b}\Big(\frac{b}{J^{2}}\Big)^{2\Delta}\Big(\frac{1}{q-1}K_{h}^{sa}W^{a}_{1-h}(\tau_{3},\tau_{4};\tau_{0}) -K^{ss}_{h}W^{s}_{1-h}(\tau_{3},\tau_{4};\tau_{0})\Big) \label{WthFtRels}
\end{cases}\,.
\end{align}

\section{\texorpdfstring{Integrals $I^{\sigma\sigma'}_{h}$ of the product  $W_{h}^{\sigma} W_{1-h}^{\sigma'}$}{Integrals I of the product W W}}
\label{IntegralsAS}
In this section, we discuss the computation of the integrals $I_h^{aa}$, $I_h^{as}$, $I_h^{sa}$, and $I_h^{ss}$ that appear in the conformal four-point function (\ref{4ptexact3}) and are defined in (\ref{IntegralsDef0}). To simplify the computation, we work in the zero-temperature limit. The finite-temperature result can then be obtained from the zero-temperature one using the map $\tau \to e^{2\pi \tau/\beta}$, i.e. from the line to the thermal circle. At zero-temperature  the integrals $I_h^{\sigma\sigma'}$ read
\begin{align}
I_{h}^{\sigma\sigma'}(\tau_{1},\tau_{2};\tau_{3},\tau_{4}) \equiv  \int_{-\infty}^{+\infty} d\tau_{0} W_{h}^{\sigma}(\tau_{1},\tau_{2}; \tau_{0}) W_{1-h}^{\sigma'}(\tau_{3},\tau_{4};\tau_{0})\,, \label{IntegralsDef}
\end{align}
where $\sigma,\sigma'=a,s$ and the right basis functions $W_{h}^{a}$ and $W_{h}^{s}$ are written in (\ref{WaWsT0}). We will show that these integrals are proportional to particular time-dependent prefactors involving the $\tau_{i}$, multiplied by functions that depend only on the cross-ratio 
\begin{align}
\chi = \frac{\tau_{12}\tau_{34}}{\tau_{13}\tau_{24}} \label{CrossR1}\,.
\end{align}
We denote these functions  by $\Psi_{h}^{\sigma\sigma'}(\chi)$ for $\sigma,\sigma'=a,s$. Indeed, we can write
\begin{equation}
\begin{aligned}
&I_{h}^{aa}(\tau_{1},\tau_{2};\tau_{3},\tau_{4}) = \Big(\frac{b}{J^{2}}\Big)^{-2\Delta} G_{c}(\tau_{12})G_{c}(\tau_{34})\Psi_{h}^{aa}(\chi)\,, \\
&I_{h}^{as} (\tau_{1},\tau_{2};\tau_{3},\tau_{4})= \Big(\frac{b}{J^{2}}\Big)^{-2\Delta} G_{c}(\tau_{12})G_{c}(\tau_{34}) \sgn(\tau_{13})\sgn(\tau_{14})\sgn(\tau_{34}) \Psi_{h}^{as}(\chi)\,, \\
&I_{h}^{sa}(\tau_{1},\tau_{2};\tau_{3},\tau_{4}) = \Big(\frac{b}{J^{2}}\Big)^{-2\Delta} G_{c}(\tau_{12})G_{c}(\tau_{34})\sgn(\tau_{12})\sgn(\tau_{13})\sgn(\tau_{23}) \Psi_{h}^{sa}(\chi)\,, \\
&I_{h}^{ss}(\tau_{1},\tau_{2};\tau_{3},\tau_{4}) = \Big(\frac{b}{J^{2}}\Big)^{-2\Delta} G_{c}(\tau_{12})G_{c}(\tau_{34})\Psi_{h}^{ss}(\chi)\,, \label{ItoPsi}
\end{aligned}
\end{equation}
where the zero temperature Green's function $G_{c}(\tau)$ is written in (\ref{GcAtT0}) and the functions $\Psi_{h}^{\sigma\sigma'}(\chi)$ are given by the following integrals 
\begin{equation}
\begin{aligned}
&\Psi_{h}^{aa}(\chi) =  \int_{-\infty}^{+\infty} d\tau_{0} \frac{|\tau_{12}|^{h}|\tau_{34}|^{1-h}}{|\tau_{10}|^{h} |\tau_{20}|^{h}|\tau_{30}|^{1-h} |\tau_{40}|^{1-h}}\,, \\
&\Psi_{h}^{as}(\chi) =  \int_{-\infty}^{+\infty} d\tau_{0} \frac{\sgn(\tau_{30})\sgn(\tau_{40})}{\sgn(\tau_{13})\sgn(\tau_{14})}\frac{|\tau_{12}|^{h}|\tau_{34}|^{1-h}}{|\tau_{10}|^{h} |\tau_{20}|^{h}|\tau_{30}|^{1-h} |\tau_{40}|^{1-h}}\,, \\
&\Psi_{h}^{sa}(\chi) =  \int_{-\infty}^{+\infty} d\tau_{0} \frac{\sgn(\tau_{10})\sgn(\tau_{20})}{\sgn(\tau_{13})\sgn(\tau_{23})} \frac{|\tau_{12}|^{h}|\tau_{34}|^{1-h}}{|\tau_{10}|^{h} |\tau_{20}|^{h}|\tau_{30}|^{1-h} |\tau_{40}|^{1-h}}\,, \\
&\Psi_{h}^{ss}(\chi) = \int_{-\infty}^{+\infty} d\tau_{0} \frac{\sgn(\tau_{10})\sgn(\tau_{20})\sgn(\tau_{30})\sgn(\tau_{40})}{\sgn(\tau_{12})\sgn(\tau_{34})} \frac{|\tau_{12}|^{h}|\tau_{34}|^{1-h}}{|\tau_{10}|^{h} |\tau_{20}|^{h}|\tau_{30}|^{1-h} |\tau_{40}|^{1-h}}\,. \label{PsiAll}
\end{aligned}
\end{equation}
These functions depend only on $\chi$ because the integrals are invariant under translations, dilatations, and inversions of  $\tau_i$, as one can easily verify.

Before we compute these integrals explicitly, we note several symmetries. First, we list how the cross-ratio transforms under exchanges of the time variables:
\begin{equation}
\begin{aligned}
&\tau_{1} \leftrightarrow \tau_{2} :\;\;   \chi \to \frac{\chi}{\chi -1}\,, \quad 
\tau_{3} \leftrightarrow \tau_{4} :\;\;   \chi \to \frac{\chi}{\chi -1}\,, \\ 
&\tau_{1} \leftrightarrow \tau_{3} :\;\;   \chi \to 1-\chi\,,\quad \ 
\tau_{2} \leftrightarrow \tau_{4} :\;\;   \chi \to 1-\chi\,. 
\end{aligned}
\end{equation}
Next we see, for example, that the integrand of $\Psi_{h}^{aa}(\chi)$ is invariant under $\tau_{1} \leftrightarrow \tau_{2}$, therefore we conclude that 
\begin{align}
&\Psi_{h}^{aa}(\chi) =  \Psi_{h}^{aa}(\chi/(\chi-1)) \,.
\end{align}
Similarly, we can find 
\begin{align}
 \Psi_{h}^{as}(\chi) =  \Psi_{h}^{as}(\chi/(\chi-1))\,, \; \Psi_{h}^{sa}(\chi) =  \Psi_{h}^{sa}(\chi/(\chi-1))\,,  \; \Psi_{h}^{ss}(\chi) =  -\Psi_h^{ss}(\chi/(\chi-1)) \,.
\end{align}
The transformation $\chi \to \chi/(\chi -1)$ maps $(-\infty,0) \leftrightarrow (0, 1)$ and $(1,2) \leftrightarrow (2,+\infty)$,
and therefore we need to compute the integrals $\Psi_{h}^{\sigma\sigma'}(\chi)$ only for $\chi \in (0,2)$.
Now setting $(\tau_{1},\tau_{2},\tau_{3}, \tau_{4})$ to $(0,\chi,1,+\infty)$, we obtain 
\begin{equation}
\begin{aligned}
&\Psi_{h}^{aa}(\chi) =  \int_{-\infty}^{+\infty} d\tau_{0} \frac{|\chi|^{h}}{|\tau_{0}|^{h} |\chi-\tau_{0}|^{h}|1-\tau_{0}|^{1-h}}\,, \\
&\Psi_{h}^{as}(\chi) =  \int_{-\infty}^{+\infty} d\tau_{0} \frac{\sgn(1-\tau_{0})|\chi|^{h}}{|\tau_{0}|^{h} |\chi-\tau_{0}|^{h}|1-\tau_{0}|^{1-h}}\,, \\
&\Psi_{h}^{sa}(\chi) =  \int_{-\infty}^{+\infty} d\tau_{0} \frac{\sgn(\chi-1)\sgn(\tau_{0})\sgn(\chi -\tau_{0})|\chi|^{h}}{|\tau_{0}|^{h} |\chi-\tau_{0}|^{h}|1-\tau_{0}|^{1-h}}\,, \\
&\Psi_{h}^{ss}(\chi) = -\int_{-\infty}^{+\infty} d\tau_{0}  \frac{\sgn(\tau_{0}) \sgn(1-\tau_{0})\sgn(\chi -\tau_{0})\sgn(\chi)|\chi|^{h}}{|\tau_{0}|^{h} |\chi-\tau_{0}|^{h}|1-\tau_{0}|^{1-h}}\,.
\end{aligned}
\end{equation}
We compute the above integrals explicitly and the final results are presented in (\ref{PsiaaToPsiss}). 

\section{\texorpdfstring{Eigenvalues and eigenvectors of $K_{c}(h)$ at integer $h$}{Eigenvalues and eigenvectors of K\_c at integer h}}
\label{AppKc}
In this appendix, we list the results for the matrix $K_{c}(h)$ and integer values of $h$. For the eigenvalues $k_{1}(h)$ and $k_{2}(h)$ in (\ref{k1k2eigs}), we find
\begin{align}
k_{1,2}(h) &= \frac{\Gamma(2\Delta-h)\Gamma(2\Delta+h-1)}{\Gamma(2\Delta+1)\Gamma(2\Delta-1)} (2\Delta - 1 \mp 1), \quad h=1,2,3,\dots\,.
\end{align}
For even positive values of $h=2,4,6,\dots$, the eigenvectors in (\ref{v1v2right}) and (\ref{w1w2left}) reduce to
\begin{equation}
\begin{aligned}
\textbf{v}_{1}(h,\theta)&=\left(  \begin{array}{c}
   1 + \frac{2\Delta \sin(2\theta)}{\sin(2\pi \Delta)} \\ 
   0 \\ 
  \end{array}\right), \quad  \textbf{v}_{2}(h,\theta)=\left(  \begin{array}{c}
   \frac{2\Delta \sin(2\theta)}{\sin(2\pi \Delta)} \\ 
   1 \\ 
  \end{array}\right)\,, \\
\textbf{w}_{1}(h,\theta)&=\frac{1}{1 + \frac{2\Delta \sin(2\theta)}{\sin(2\pi \Delta)}}\left(  \begin{array}{c}
   1 \\ 
   -\frac{2\Delta \sin(2\theta)}{\sin(2\pi \Delta)} \\ 
  \end{array}\right), \quad  \textbf{w}_{2}(h,\theta)=\left(  \begin{array}{c}
   0 \\ 
   1 \\ 
  \end{array}\right) \,. \label{vweven}
\end{aligned}
\end{equation}
For odd positive values of $h=1,3,5,\dots$, they are
\begin{equation}
\begin{aligned}
\textbf{v}_{1}(h,\theta)&=\left(  \begin{array}{c}
   1 \\ 
 \frac{2(1-\Delta) \sin(2\theta)}{\sin(2\pi \Delta)}  \\ 
  \end{array}\right), \quad  \textbf{v}_{2}(h,\theta)=\left(  \begin{array}{c}
   0 \\ 
   1+\frac{2(1-\Delta) \sin(2\theta)}{\sin(2\pi \Delta)} \\ 
  \end{array}\right) \,, \\
\textbf{w}_{1}(h,\theta)&=\left(  \begin{array}{c}
   1 \\ 
   0\\ 
  \end{array}\right), \quad  \textbf{w}_{2}(h,\theta)=\frac{1}{1 + \frac{2(1-\Delta) \sin(2\theta)}{\sin(2\pi \Delta)}}\left(  \begin{array}{c}
   -\frac{2(1-\Delta) \sin(2\theta)}{\sin(2\pi \Delta)}  \\ 
   1 \\ 
  \end{array}\right)\,. \label{vwodd}
\end{aligned}
\end{equation}
Using this, we can write the matrix $K_{c}(h)$ in the form 
\begin{equation}
\begin{aligned}
&K_{c}(h) = \frac{2\Gamma(2\Delta-h)\Gamma(2\Delta+h-1)}{\Gamma(2\Delta+1)\Gamma(2\Delta-1)} \left(  \begin{array}{cc}
    \Delta-1 & \frac{2\Delta \sin(2\theta)}{\sin(2\pi \Delta)} \\ 
    0 & \Delta \\ 
  \end{array}\right)\,, \quad h= 2,4,6,\dots\,, \\
&K_{c}(h) = \frac{2\Gamma(2\Delta-h)\Gamma(2\Delta+h-1)}{\Gamma(2\Delta+1)\Gamma(2\Delta-1)} \left(  \begin{array}{cc}
    \Delta-1 & 0 \\ 
    \frac{2(\Delta-1) \sin(2\theta)}{\sin(2\pi \Delta)} & \Delta \\ 
  \end{array}\right)\,, \quad h= 1,3,5,\dots\,\,. 
\end{aligned}
\end{equation}

\section{Conformal generators of descendants}
\label{AppCompOPE}
Starting from the OPE \eqref{ccOPE} and using the conformal 1 and 2-point functions, we get
\begin{align}
    &\frac{1}{N}\langle \psi^{i}(\tau_{1})\psibar_{i}(\tau_{2}) \mathcal{O}_{h}(\tau_0)\rangle= \frac{G_{c}(\tau_{12})}{|\tau_{12}|^{-h}} \left[  c_{h}^{a}C^{a}_{h}(\tau_{12},\partial_2)+  c_{h}^{s} \sgn(\tau_{12}) C^{s}_{h}(\tau_{12},\partial_2)\right] \frac{1}{|\tau_{20}|^{2h}}\,.
\end{align}
Matching this result with the one-dimensional conformal 3-point function \eqref{eq_3pt_function}, we arrive at the relations:
\begin{align}
    C^{a}_{h}(\tau_{12},\partial_2) \frac{1}{|\tau_{20}|^{2h}}= \frac{1}{|\tau_{10}|^{h}|\tau_{20}|^{h}}\,, \quad 
    C^{s}_{h}(\tau_{12},\partial_2) \frac{1}{|\tau_{20}|^{2h}}= \frac{\sgn(\tau_{10})\sgn(\tau_{20})}{|\tau_{10}|^{h}|\tau_{20}|^{h}}\,.
    \label{eq_gen_equations}
\end{align}
In order to calculate $C_{h}^{s}(\tau_{12},\partial_2)$, we rewrite the right-hand side as
\begin{equation}
\begin{aligned}
    C^{s}_{h}(\tau_{12},\partial_2)  \frac{1}{|\tau_{20}|^{2h}} &= \frac{\sgn(\tau_{10})\sgn(\tau_{20})}{|\tau_{10}|^{h}|\tau_{20}|^{h}}=\frac{(\tau_{12}+\tau_{20})\tau_{20}}{|\tau_{10}|^{h+1}|\tau_{20}|^{h+1}} \\
    &=\frac{1}{|\tau_{10}|^{h+1}|\tau_{20}|^{h-1}}+\frac{\tau_{12}\tau_{20}}{|\tau_{10}|^{h+1}|\tau_{20}|^{h+1}}\\
    &= \left(1-\frac{1}{h-1}\tau_{12}\partial_{2}\right)\frac{1}{|\tau_{10}|^{h+1}|\tau_{20}|^{h-1}}\\
    &= \left(1-\frac{1}{h-1}\tau_{12}\partial_{2}\right) \tilde{C}_{h}^s(\tau_{12},\partial_2) \frac{1}{|\tau_{20}|^{2h}}\,.
    \label{eq_CsDs}
\end{aligned}
\end{equation}
Let us now calculate $\tilde{C}_{h}^s(\tau_{12},\partial_2)$ by making the ansatz
\begin{align}
    \tilde{C}_{h}^s(\tau_{12},\partial_2)= \sum_{k=0}^{\infty}d^{s}_{h,k}(\tau_{12})^{k}\partial_{2}^{k}\,,
\end{align}
equating the last two lines of \eqref{eq_CsDs}, we find that $d^{s}_{h,k}=\frac{(h+1)_{k}}{(2h)_{k}k!}$, where $(\ell)_{k}\equiv\frac{\Gamma(\ell+k)}{\Gamma(\ell)}$ is the Pochhammer symbol. Using the same ansatz, we find that the coefficients of $C_{h}^{a}$ are $d^{a}_{h,k}=\frac{(h)_{k}}{(2h)_{k}k!}$. So, we conclude that the generators of the descendants can be written in the following short form:
\begin{equation}
\begin{aligned}
    C_{h}^{a}(\tau_{12},\partial_2)&=\,_1F_1(h,2h;\tau_{12}\partial_2)\,, \\
    C_{h}^{s}(\tau_{12},\partial_2)&= \left(1-\frac{1}{h-1}\tau_{12}\partial_{2}\right)\,_1F_1(h+1,2h;\tau_{12}\partial_2)\,,
\end{aligned}
\end{equation}
where it is assumed that derivative $\partial_{2}$ always acts first, so we have $(\tau_{12})^{k}\partial_{2}^{k}$ instead of $(\tau_{12}\partial_{2})^{k}$. The above formulas and the formulas which we derive below, have appeared many times in the literature for various spacetime dimensions, see, e.g. \cite{Ferrara:1972kab, Petkou:1994ad, Dolan:2000uw, Dolan:2000ut, Fortin:2020zxw}. 

First, we derive the formula 
\begin{align}
C^{a}_{h}(\tau_{12},\partial_{2}) \frac{1}{|\tau_{23}|^{h+h_{12}}|\tau_{24}|^{h-h_{12}}}  =\frac{1}{|\tau_{12}|^{h}|\tau_{34}|^{h}} \left|\frac{\tau_{14}}{\tau_{13}}\right|^{h_{12}}
 |\chi|^{h}\,_{2}F_{1}(h, h+h_{12},2h,  \chi)\,. \label{DaF}
\end{align} 
For this, we follow closely Appendix D of \cite{Dolan:2000uw}. Using the Feynman parametrization
  \begin{align}
&\frac{1}{A^{a}B^{b}}= \frac{1}{B(a,b)}\int_{0}^{1}d\alpha \frac{\alpha^{a-1}(1-\alpha)^{b-1} }{(\alpha A + (1-\alpha)B)^{a+b}} \,,
\quad B(a,b)\equiv\frac{\Gamma(a)\Gamma(b)}{\Gamma(a+b)}\,, 
\label{Feynmparam}
\end{align} 
and assuming  that $\tau_{23},\tau_{24}>0$ as well as $\tau_{12}>0$, we find 
\begin{equation}
\begin{aligned}
&C^{a}_{h}(\tau_{12},\partial_{2}) \frac{1}{|\tau_{23}|^{h+h_{12}}|\tau_{24}|^{h-h_{12}}}=  \\
&\qquad\qquad =\frac{1 }{B(h+h_{12},h-h_{12})}\int_{0}^{1}d\alpha \frac{\alpha^{h+h_{12}-1}(1-\alpha)^{h-h_{12}-1}}{(\tau_{14}-\alpha \tau_{34})^{h}(\tau_{24}-\alpha\tau_{34})^{h}}\,,
\end{aligned}
\end{equation}
where after applying the Feynman parametrization we used (\ref{eq_gen_equations}).
Then we integrate over $\alpha$ using the formula 
\begin{align}
\int_{0}^{1}d\alpha \frac{\alpha^{e-1}(1-\alpha)^{f-1} }{(1-\alpha x)^{a}(1-\alpha y)^{b}} = \frac{B(e,f)}{(1-x)^{e}}\,_{2}F_{1}\Big(e,b;a+b; \frac{y-x}{1-x}\Big)\,, \label{Int2F1}
\end{align} 
where $a+b= e+f$, and obtain (\ref{DaF}). The formula (\ref{Int2F1}) can be derived by applying the Feynman parametrization (\ref{Feynmparam}) to the denominator and then expanding it in powers of $(y-x)/(1-\alpha x)$. After integrating over both Feynman parameters, one obtains the final result.

Next, similarly to (\ref{DaF}), we derive the formula
\begin{align}
&\tilde{C}_{h}^{s}(\tau_{12},\partial_2)\frac{1}{|\tau_{23}|^{h+h_{12}}|\tau_{24}|^{h-h_{12}}} =
\frac{|\tau_{14}|^{h_{12}-1}|\tau_{24}|}{|\tau_{12}|^{h}|\tau_{34}|^{h}|\tau_{13}|^{h_{12}}} |\chi|^{h}\,_2F_1\left(h-1,h+h_{12};2h;\chi\right)\label{DtS}
\end{align} 
For this, we assume  that $\tau_{23},\tau_{24}>0$ as well as $\tau_{12}>0$, and find
\begin{equation}
\begin{aligned}
&\tilde{C}^{s}_{h}(\tau_{12},\partial_{2}) \frac{1}{|\tau_{23}|^{h+h_{12}}|\tau_{24}|^{h-h_{12}}}=  \\
&\qquad\qquad =\frac{1 }{B(h+h_{12},h-h_{12})}\int_{0}^{1}d\alpha \frac{\alpha^{h+h_{12}-1}(1-\alpha)^{h-h_{12}-1}}{(\tau_{14}-\alpha \tau_{34})^{h+1}(\tau_{24}-\alpha\tau_{34})^{h-1}}\,.
\end{aligned}
\end{equation}
Applying the formula (\ref{Int2F1}) we arrive at (\ref{DtS}).

Finally, by using (\ref{DaF}) and (\ref{DtS}) several times, we can obtain the following formulas: 
\begin{equation}
\begin{aligned}
C_{h}^{a}(\tau_{12},\partial_2) C_{h}^{a}(\tau_{34},\partial_4)\frac{1}{|\tau_{24}|^{2h}}&=\frac{1}{|\tau_{12}\tau_{34}|^h} |\chi|^{h}\,_2F_1(h,h,2h;\chi)\,,\\
C_{h}^{a}(\tau_{12},\partial_2) C_{h}^{s}(\tau_{34},\partial_4)\frac{1}{|\tau_{24}|^{2h}}&= \frac{\sgn(\tau_{13})\sgn(\tau_{14})}{|\tau_{12}\tau_{34}|^h} |\chi|^h\,_2F_1(h,h;2h;\chi)\,, \\ 
C_{h}^{s}(\tau_{12},\partial_2) C_{h}^{a}(\tau_{34},\partial_4)\frac{1}{|\tau_{24}|^{2h}}&= \frac{\sgn(\tau_{13})\sgn(\tau_{23})}{|\tau_{12}\tau_{34}|^h} |\chi|^h\,_2F_1(h,h;2h;\chi)\,,\\
C_{h}^{s}(\tau_{12},\partial_2)C_{h}^{s}(\tau_{34},\partial_4)\frac{1}{|\tau_{24}|^{2h}}&= \frac{\sgn(\tau_{13})\sgn(\tau_{24})}{|\tau_{12}\tau_{34}|^h} |\chi|^{h}\,_2F_1(h,h;2h;\chi)\,. \label{DFall}
\end{aligned}
\end{equation}

\section{Four-point function from the shadow formalism}\label{app:shadowformalism}
We provide in this appendix an alternate derivation of \eqref{conformal_4ptMain} using the shadow formalism.
We use some standard facts from the shadow formalism \cite{Ferrara:1972xe, Ferrara:1972ay, Ferrara:1972uq, Ferrara:1972kab, Simmons-Duffin:2012juh, Fraser-Taliente:2025udk} and follow closely \cite{Simmons-Duffin:2012juh}.
Taking $\mathcal{O}$ to be a real scalar operator of dimension $\Delta$, with a unit-normalized two-point function, we define 
a shadow transform $\mathbf{S}[\mathcal{O}](x) = \tilde{\mathcal{O}}(x)$ to be
\begin{equation}
     \tilde{\mathcal{O}}(x) \equiv \int d^d y  \frac{\mathcal{O}(y)}{|x-y|^{2(d-\Delta)}}\,,
\end{equation}
then using that $\langle \mathcal{O}(x) \mathcal{O}(y)\rangle = |x-y|^{-2\Delta}$ we find 
\begin{equation}
\langle {\tilde{\mathcal{O}}(x) \tilde{\mathcal{O}}(y)}\rangle =\frac{\mathcal{N}_{\mathcal{O}} }{|x-y|^{2(d-\Delta)}}, \quad 
\mathcal{N}_{\mathcal{O}} = \frac{\pi^{d}\Gamma(\Delta-d/2)\Gamma(d/2-\Delta)}{\Gamma(\Delta)\Gamma(d-\Delta)}\,,
\end{equation}
and also $\mathbf{S}^2[\mathcal{O}](x) = \mathcal{N}_{\mathcal{O}} \mathcal{O}(x)$.
Then, the following object \cite[(2.33)]{Simmons-Duffin:2012juh}
\begin{equation}\label{eq:absOdef}
|\mathcal{O}| = \frac{1}{\mathcal{N}_{\mathcal{O}}} \int d^d x |\mathcal{O}(x)\rangle \langle\tilde{\mathcal{O}}(x)| + \text{monodromy} 
\end{equation}
is \textit{almost} a unit-normalised projector onto the conformal multiplet of $\mathcal{O}$, where \enquote{$+\,\textrm{monodromy}$} indicates that we just need to supplement its insertion by an appropriate monodromy projection.
This projection serves to remove the contribution of the shadow block, which transforms differently under the monodromy.
In particular, we define $|\mathcal{O}|$ by \cite[(2.34)]{Simmons-Duffin:2012juh}
\begin{equation}
    \langle \phi_1 \cdots \phi_m |\mathcal{O}|\phi_{m+1} \cdots \phi_n\rangle = \frac{1}{\mathcal{N}_{\mathcal{O}}} \int d^d{x} \langle{\phi_1 \cdots \phi_m \mathcal{O}(x)}\rangle\langle{\tilde{\mathcal{O}}(x)\phi_{m+1}\cdots \phi_n}\rangle\Big\rvert_{M=e^{2\pi i \varphi}}\,,
\end{equation}
where $M$ is the monodromy map which is convenient to implement in the embedding space \cite[(2.9)]{Simmons-Duffin:2012juh}: its action on the embedding space coordinates $X_{ij}$ is $X_{ij} \mapsto e^{4\pi i} X_{ij}$ for $i,j \leq m$, and leaves the other $X_{ij}$ invariant.
The $\rvert_{M=e^{2\pi i \varphi}}$ indicates that we project to the eigenspace of $M$ on this correlator with eigenvalue $e^{2\pi i \varphi}$.
Consistency with the OPE makes it easy to determine that we require $\varphi \equiv \Delta - \sum_{i\le m} \Delta_i$, where $\Delta_i$ is the scaling dimension of $\phi_i$.

Hence, in the shadow formalism taking $d=1$, one can insert a complete set of states into a correlation function using \footnote{We note that the normalization constant here differs from (\ref{NormConst}) by an overall factor of $1/2$.}
\begin{align}\label{eq:identityDecomp}
\sum_{h= h_{m},\mathbb{I}} \frac{1}{N\mathcal{N}_{h}} \int_{-\infty}^{+\infty} d\tau_{0} \, | \mathcal{O}_{h} (\tau_{0}) \rangle \langle \tilde{\mathcal{O}}_{h}(\tau_{0})| = \textrm{Identity}\,, \quad \frac{1}{\mathcal{N}_{h}} = \frac{(h-1/2)}{\pi \tan (\pi  h)}\,,
\end{align}
where $\mathcal{O}_{h_{m}}$ are  all primary operators (not including their shadows), which appear in the OPE of $\psi^{i}\times \bar{\psi}_i$.
We explicitly distinguish the identity operator $\mathbb{I}$ from the rest of the operators, since we will be subtracting its contribution shortly.  
Additionally, \eqref{eq:identityDecomp} contains a factor of $1/N$ compared to \eqref{eq:absOdef} to account for the fact that, in our convention in \eqref{ccOPE}, the $\mathcal{O}_h$ are not unit-normalized:
\begin{equation}
    \langle \mathcal{O}_h(\tau) \mathcal{O}_{h'}(0)\rangle = \frac{N\delta_{hh'}}{|\tau|^{2h}}.
\end{equation}

Inserting \eqref{eq:identityDecomp} into the connected four-point function $\mathcal{F}$ as defined by \eqref{eq:Fdef}, and dropping terms that are subleading in $N$, we find
\begin{equation}
\begin{aligned}
&\mathcal{F}(\tau_{1},\tau_{2};\tau_{3},\tau_{4}) = \tfrac{1}{N}\langle \psi^{i}(\tau_{1}) \psibar_{i}(\tau_{2})\psi^{j}(\tau_{3})\psibar_{j}(\tau_{4})\rangle  - N G(\tau_{12})G(\tau_{34})\\
&\quad =  \sum_{h=h_{m}} \frac{1}{N^2\mathcal{N}_{h}} \int d\tau_{0}\langle\psi^{i}(\tau_{1}) \psibar_{i}(\tau_{2}) \mathcal{O}_{h}(\tau_{0})\rangle   \langle \tilde{\mathcal{O}}_{h}(\tau_{0})
\psi^{j}(\tau_{3})\psibar_{j}(\tau_{4})\rangle \Big\rvert_{M=e^{2\pi i \varphi}} \\
&\quad =  \sum_{h=h_{m}} \frac{1}{N^2\mathcal{N}_{h}} \int d\tau_{0}\,d\tau_{5 }\langle \psi^{i}(\tau_{1}) \psibar_{i}(\tau_{2}) \mathcal{O}_{h}(\tau_{0})\rangle  \frac{1}{|\tau_{05}|^{2(1-h)}} \langle \mathcal{O}_{h}(\tau_{5})\psi^{j}(\tau_{3})\psibar_{j}(\tau_{4})\rangle  \Big\rvert_{M=e^{2\pi i \varphi}}. \label{eq:Fshadowblocks}
\end{aligned}
\end{equation}
Observe that the sum no longer includes the contribution of the identity operator, which was explicitly subtracted by the leading $N$ term. To proceed, we write the three-point functions \eqref{eq_3pt_function} in the form
\begin{align}
\frac{1}{N}\langle \psi^i(\tau_{1}) \psibar_i(\tau_{2})\mathcal{O}_{h}(\tau_{0})\rangle = \Big(\frac{b}{J^2}\Big)^{\Delta} \Big(c_{h}^{a} W_{h}^{a}(\tau_{1},\tau_{2};\tau_{0}) + c_{h}^{s} W_{h}^{s}(\tau_{1},\tau_{2};\tau_{0})\Big)\,, \label{3ptasPhi}
\end{align}
for $W_{h}^a$ and $W_{h}^s$ defined in \eqref{WhaWhs}. Therefore,
\eqref{eq:Fshadowblocks} becomes
\begin{equation}
\begin{aligned}
\mathcal{F}_{c}(\tau_{1},\tau_{2};\tau_{3},\tau_{4}) =&\sum_{\sigma,\sigma'=a,s} \sum_{h= h_{m}} \Big(\frac{b}{J^2}\Big)^{2\Delta}\frac{c_{h}^{\sigma} c_{h}^{\sigma'}}{\mathcal{N}_{h}} \\
&\times\int_{-\infty}^{+\infty} d\tau_{0}d \tau_5 \, W_{h}^{\sigma}(\tau_{1},\tau_{2}; \tau_{0}) \frac{1}{|\tau_{05}|^{2(1-h)}} W_{h}^{\sigma'}(\tau_{3},\tau_{4}; \tau_{5})\Big\rvert_{M=e^{2\pi i \varphi}} \,,
\end{aligned}
\end{equation}
where the factors of $N$ have now cancelled. 
We were free to move around the $\mathcal{O}_{h}$ in the second three-point function because $\mathcal{O}_h$ is a bosonic operator and the correlators are the usual ordered products.
We take the integral over $\tau_{5}$ using the star-triangle identities (\ref{st_tr_ids}) and find 
\begin{equation}
 \int_{-\infty}^{+\infty} d \tau_5 \, \frac{1}{|\tau_{05}|^{2(1-h)}} W_{h}^{\sigma'}(\tau_{3},\tau_{4}; \tau_{5})  = \kappa^{\sigma'}_{h} W_{1-h}^{\sigma'}(\tau_{3},\tau_{4}; \tau_{0})\,,
\end{equation}
where $\kappa^{a}_{h}=b_{h/2,h/2}$ and $\kappa^{s}_{h}=f_{h/2,h/2}$. 
The remaining integral over $\tau_{0}$ gives the functions $I^{\sigma\sigma'}_{h}$,  computed in Appendix \ref{IntegralsAS}; and thus, the four-point function takes the form
\begin{equation}
\begin{aligned}
    \mathcal{F}_{c}(\tau_{1},\tau_{2};\tau_{3},\tau_{4})&= \sum_{\sigma, \sigma' =a,s } \sum_{h=h_{m}}\Big(\frac{b}{J^2}\Big)^{2\Delta}c_{h}^{\sigma} c_{h}^{\sigma'} \frac{ \kappa^{\sigma'}_{h}}{\mathcal{N}_{h}} I^{\sigma \sigma'}_{h}(\tau_{1},\tau_{2};\tau_{3},\tau_{4})  \Big\rvert_{M=e^{2\pi i \varphi}}\,.
\end{aligned}
\end{equation}
Using the expressions in (\ref{ItoPsi}) for  $I^{\sigma \sigma'}_{h}$,  we find
\begin{equation}
\begin{aligned}
\frac{\mathcal{F}_{c}(\tau_{1},\tau_{2};\tau_{3},\tau_{4})}{G_{c}(\tau_{12})G_{c}(\tau_{34})}=&  \sum_{h=h_{m}}\Big((c_{h}^{a})^{2}\frac{ \kappa^{a}_{h}}{\mathcal{N}_{h}}\Psi_{h}^{aa}(\chi)+c_{h}^{a}c_{h}^{s}\sgn(\tau_{13}\tau_{14}\tau_{34})\frac{ \kappa^{s}_{h}}{\mathcal{N}_{h}}\Psi_{h}^{as}(\chi)  \\ 
&\quad+c_{h}^{s}c_{h}^{a}\sgn(\tau_{12}\tau_{13}\tau_{23})\frac{ \kappa^{a}_{h}}{\mathcal{N}_{h}}\Psi_{h}^{sa}(\chi)+
(c_{h}^{s})^{2}\frac{ \kappa^{s}_{h}}{\mathcal{N}_{h}}\Psi_{h}^{ss}(\chi)\Big)\Big\rvert_{M}\,.
\end{aligned}
\end{equation}
The functions $\Psi_{h}^{\sigma\sigma'}(\chi)$ are conformal partial waves \cite{Mazac:2018qmi}, and for $0<\chi<1$ they are linear combinations of a conformal block and its shadow.
Finally, using that
\begin{equation}
    \frac{\kappa^{a}_{h}}{\mathcal{N}_{h}} = \frac{\tan \left(\frac{\pi  h}{2}\right)}{\tan(\pi h)}\frac{\Gamma(2h)}{\Gamma(h)^2}, \quad \frac{\kappa^{s}_{h}}{\mathcal{N}_{h}} = \frac{\cot \left(\frac{\pi  h}{2}\right)}{\tan(\pi h)}\frac{\Gamma(2h)}{\Gamma(h)^2} \,,
\end{equation}
together with the explicit expressions for $\Psi^{\sigma\sigma'}_{h}(\chi)$ in \eqref{PsiaaToPsiss} for $0<\chi <1$, where the OPE is valid, and using that the monodromy projection removes the shadow blocks, we arrive at the result in (\ref{conformal_4ptMain}).

\bibliography{csykcft.bib}

\end{document}